\documentclass{article}
\usepackage{amsmath,amssymb}
\usepackage{epsfig,multirow}
\usepackage{array}
\usepackage[a4paper,top=3cm,bottom=2cm,left=3cm,right=3cm,marginparwidth=1.75cm]{geometry}

\title{Large-scale anisotropies of extragalactic cosmic rays \\below the ankle}
\author{S. Mollerach$^1$, E. Roulet$^1$ and  O. Taborda$^2$\\$^1$ Centro At\'omico Bariloche, Comisi\'on Nacional de Energ\'\i a At\'omica\\ 
Consejo Nacional de Investigaciones Cient\'\i ficas y T\'ecnicas (CONICET)\\ 
Av. Bustillo 9500, R8402AGP, Bariloche, Argentina\\
$^2$ Universidad Tecnol\'ogica de Pereira, Complejo Educativo La Julita \\ Pereira, Risaralda, Colombia}
\date{}

\begin{document}

\maketitle
\begin{abstract}
We study the anisotropies on large angular scales which can be present in the flux of cosmic rays reaching the Earth from a population of extragalactic sources, focusing on the energy range between the second knee and the ankle. In this energy range the particles are significantly affected by the Galactic magnetic field, which then plays a relevant role in shaping the expected anisotropies. The Galactic magnetic field deflects  the  cosmic-ray trajectories and thus modifies the anisotropies present outside the halo of the Galaxy, in particular the dipolar one associated with the translational motion of the observer (Compton-Getting effect). Also, due to the Galactic rotation, in the reference frame of an observer at Earth there is an electric component of the Galactic field that produces a small change in the  particles' momentum. This acceleration depends on the cosmic-ray arrival direction and it hence induces anisotropies in the flux observed in a given energy range. We analyse the expected amplitude and phase of the resulting dipolar component of the flux and discuss the possibility to explain via these effects the change in the phase of the right-ascension distribution which is observed at energies around 1~EeV.
\end{abstract}

\section{Introduction}

The study of the anisotropies in the arrival directions is expected to play a fundamental role, together with the spectrum and composition measurements, for understanding  the origin and the propagation of high-energy cosmic rays (CRs).  For energies between $10^{10}$~eV and $10^{20}$~eV
the overall CR spectrum has a shape which is close to a power law, decreasing approximately as $E^{-3}$. Its
 main features are the knee, which indicates the steepening at around 3 to 4~PeV, the second knee, corresponding to a further steepening  at about 100~PeV \cite{ICspec,TAspec}, there is then a hardening at  the ankle around 5~EeV, followed by a strong falloff starting at around 50~EeV \cite{Augerspec}. 
 
 A widely accepted explanation for the knee features, also in agreement with the  composition trend   from lighter to heavier nuclei which is measured in the knee to second-knee energy range, is that the spectrum of Galactic protons starts to fade at the knee while that of the Galactic irons starts to fade at an energy 26 times higher, as a consequence of a rigidity dependent phenomenon.  Above the second knee and  up to about 2~EeV the composition becomes progressively lighter  for increasing energies \cite{Augercomp}. There are good reasons to believe that this population of light cosmic rays, showing up between the second knee and the ankle, has an extragalactic origin. On the one hand, no Galactic source is expected to be able to accelerate light elements up to those energies and, on the other hand, Galactic protons would give rise to exceedingly large anisotropies in the distribution of arrival directions, in disagreement with observations \cite{abreu13,lsa20}. Indeed, no significant anisotropies have been measured between the second knee and the ankle, with the upper bounds obtained on the equatorial component of the dipole amplitude being at the level of 1 to 2\% in this energy range. It is anyway worth to note that  a change in the right-ascension phase of the reconstructed dipole is observed, going from directions close to the Galactic center one,  $\alpha \simeq 270^\circ$, at energies below about 1~EeV \cite{LSAIC,LSAKG}, to  the nearly opposite direction, $\alpha \simeq 90^\circ$, above 4~EeV \cite{LSAscience,lsa20}. Regarding the hardening observed at the ankle, it is probably associated with the emergence of a new extragalactic component of powerful and closeby sources, while the suppression at the highest energies is probably associated with a combination of the effects of the maximum achievable acceleration at the sources and the attenuation effects that the CR suffer during their propagation through  radiation fields.

Extragalactic cosmic rays in the energy range between the second knee and about $2$~EeV can  travel large distances without interacting significantly with the extragalactic background radiation, with their energies being just  redshifted by the expansion of the Universe. Since they can reach the Earth from cosmological distances, the distribution of their directions of arrival  to the Milky Way is expected to be  close to isotropic, although some anisotropies could arise from inhomogeneities in the distribution of their sources, combined with the effects of the diffusion in the intergalactic magnetic fields. Even in the case of a perfectly isotropic extragalactic distribution, a dipolar anisotropy would be observed from  Earth if there is a relative motion between the observer and the reference frame in which the CR distribution looks isotropic. Furthermore, after reaching the Galaxy, the cosmic rays have to traverse the Galactic magnetic field before arriving to  Earth. This modifies the arrival direction distribution in two ways. On one side, the trajectories are deflected by the magnetic field, and thus the angular distribution of the flux reaching the Galaxy gets distorted upon arrival to  Earth. In particular, this will change the direction and amplitude of a dipolar anisotropy present outside the Galaxy as seen from  Earth, and will also give rise to higher order multipoles. Note that this effect alone cannot  generate an anisotropy if the original flux is isotropic. On the other side,  in the reference frame of an observer at  Earth  there will be,  as a consequence of the Galactic rotation, an electric component of the Galactic  field, and hence the CRs arriving from extragalactic space will suffer a small change in their momentum as they travel through the Galaxy \cite{hmr10}. This acceleration is different for trajectories reaching the Earth from different directions and it hence  induces anisotropies in the flux observed in any given energy range, even in the case in which the original flux outside the Galaxy were isotropic. The detailed study of these different anisotropies will be the subject of the present work, extending the study of ref.~\cite{hmr10} to lower energies, considering updated and more complete magnetic field models and confronting with recent experimental results.

\section{The Galactic magnetic field and its effects on the flux of extragalactic cosmic rays}

Let us  consider the  case in which there is a reference frame for which the extragalactic CRs are isotropic  outside the Galaxy, with a flux $\Phi'(p)$ depending only on the particle's momentum $p$, but independent from its  direction $\hat u$ (where $\vec p=p\hat u$). The flux of particles arriving to  Earth with momentum $p_0$ from the direction $\hat u_0$, $\Phi (p_0,\hat u_0)$, considering an energy spectrum proportional to $E^{-\gamma}$, is related with that outside the Galaxy 
through \cite{hmr10}
\begin{equation}
\Phi (p_0,\hat u_0) \simeq \Phi'(p_0)\left[1+ (\gamma + 2)\frac{\hat u_h \cdot \vec V_{\rm iso}}{c}-(\gamma+2)\frac{p_h - p_0}{p_0} 
 \right],
\label{phiearth}
\end{equation}
 as a function of the associated directions ${\hat u_h}$ and the momenta $p_h$ that the particles  had before entering the Galactic magnetic field region (i.e. just outside the Galactic halo), with $c$ being the speed of light.
 
The second term in the parenthesis in Eq.\,(\ref{phiearth}) corresponds to the well-known Compton-Getting effect \cite{CG}, that leads to a dipolar anisotropy with an amplitude proportional to the relative velocity $V_{\rm iso}$ of the observer with respect to the frame in which CRs are isotropic. If this frame coincides with the rest frame of the cosmic microwave background (CMB), the solar system relative velocity will be $V_{\rm iso}\simeq V_{\rm CMB} \simeq 371 $\,km\,s$^{-1}$, in the direction with Galactic coordinates $(l,b) \simeq (264.4^{\circ},48.4^{\circ})$  \cite{fi96}.
Ignoring the deflections induced by the Galactic magnetic field (i.e. if $\hat u_0=\hat u_h$), this motion would induce a dipole with amplitude $d \simeq 0.006$ for a spectral index $\gamma \simeq 3$ \cite{KS}. However, the deflections of the particles' trajectories  lead  to a non-trivial mapping between the directions of incidence at the halo ${\hat u_h}$ and the arrival directions at Earth ${\hat u_0}$, distorting the dipolar pattern.

The third term within the parentheses  in Eq.\,(\ref{phiearth}) corresponds to the effect of the change of momentum  along the trajectory, which has to be evaluated through the integral \cite{hmr10}
\begin{equation}
\frac{p_h - p_0}{p_0} \simeq \frac{0.9\ {\rm EeV}}{E/Z}
\int \frac{{\rm d}\vec l}{\rm kpc} \cdot \left[\frac{(\vec V - \vec V_\odot)}{c} 
\times \frac{\vec B}{\mu {\rm G}} \right].
\label{phalo}
\end{equation}
Note that due to the high conductivity of the Galactic 
medium, we can consider that  in the reference frame moving locally with the plasma there are only magnetic fields present, with no electric fields. This ionised gas  moves  with a velocity  $\vec V$ which is approximately equal to the one of the rotation curve of the Galaxy.
To an observer moving with the solar system velocity $\vec V_\odot$, CRs traveling through far away regions of the Galaxy will then also experience an electric force coming 
from the relative motion of the medium in which the field is purely magnetic, given by $q \vec E = (q/c) \Gamma_{\Delta V} (\vec V - \vec V_\odot) \times \vec B'$, with $\Gamma_{\Delta V} \simeq 1$ being the Lorentz factor associated with the relative velocity between the local system and the observer, $\Delta\vec V \equiv \vec V - \vec V_\odot$, and $\vec B'$ is the magnetic field in the local frame (since $\Gamma_{\Delta V} \simeq 1$, we can also consider that $\vec B \simeq \vec B'$). 
To estimate $|\vec V - \vec V_\odot|$ we will just consider the plasma to be moving azimuthally with a flat rotation curve at 220\,km\,s$^{-1}$ in a clockwise direction as seen from the North.
Let us note that since the values of $|\vec V - \vec V_\odot|$ are roughly of the same order of magnitude as $V_{\rm CMB}$, for a magnetic field of strength $\sim \mu$G and a propagation length of few kpc within the Galaxy,  for values of $E/Z$ smaller than about 1~EeV the momentum change effect could give a contribution of similar or larger size than the Compton-Getting one. Then, this effect is expected to play a relevant role in shaping the anisotropies of extragalactic cosmic rays in the energy range from the second knee to the ankle.

The resulting anisotropies in the arrival directions depend on the precise model of the Galactic magnetic field, which is however not well known. For definiteness, we will adopt as  reference  model the one by Jansson and Farrar (JF12),
which is constrained by  Faraday rotation measures and the synchrotron emission maps from the WMAP satellite \cite{JF12reg,JF12ran}. In order to exemplify the dependence of the results on the magnetic field model adopted, we will also consider the modifications to this model proposed in \cite{planck16} (JF+Planck), which provide a better fit to the synchrotron and dust emission maps determined with the Planck satellite. In these models there is a coherent regular field, a random striated (ordered) field and a random isotropic field. The coherent field is described by a superposition of disk, toroidal halo and “X-field” components \cite{JF12reg}. The random striated field is aligned with the regular field, has vanishing mean amplitude and root-mean-square (rms)  strength $B_{\rm str}^2 = \beta |B_{\rm reg}|^2$. The random isotropic field consists of a superposition of  disk and halo components, with rms strength $B_{\rm rand}^2 = B_{\rm disk}^2 + B_{\rm halo}^2$.  For the coherence length of the random fields we will consider a reference value of $l_{\rm c}=50$\,pc \cite{ha08,beck16}.
 The effect of the striated field is taken into account by considering $\vec{B}_{\rm str}=b\vec{B}_{\rm reg}$,  with the parameter $b$ being changed randomly after the particles travel a distance $l_{\rm c}$, sampling it from a Gaussian distribution with variance $\beta$. Note that for the striated component the JF12 model adopts $\beta=1.36$ while the JF+Planck model considers $\beta=10$.  The effect of the random isotropic field is taken into account including in the Lorentz equation a stochastic deflection term, as originally introduced in ref.~\cite{achterberg}. This treatment is expected to give a good estimate of the mean flux distribution over different realizations of the random field, as well as of the resulting anisotropies on large angular scales, while for the rigidities that are significantly affected by the random field the small scale details   depend on the actual realization of the random field deflections that one may consider.
Since the different components have different associated uncertainties,
we will separately display the effects resulting from just the regular magnetic field component, the regular plus striated components and the complete one including also the isotropic random contribution, for both magnetic field models.

\subsection{Anisotropies due to the acceleration by the electric force}
\label{acceleration}

\begin{figure}[]
    \centering
    \includegraphics[width=0.49\textwidth]{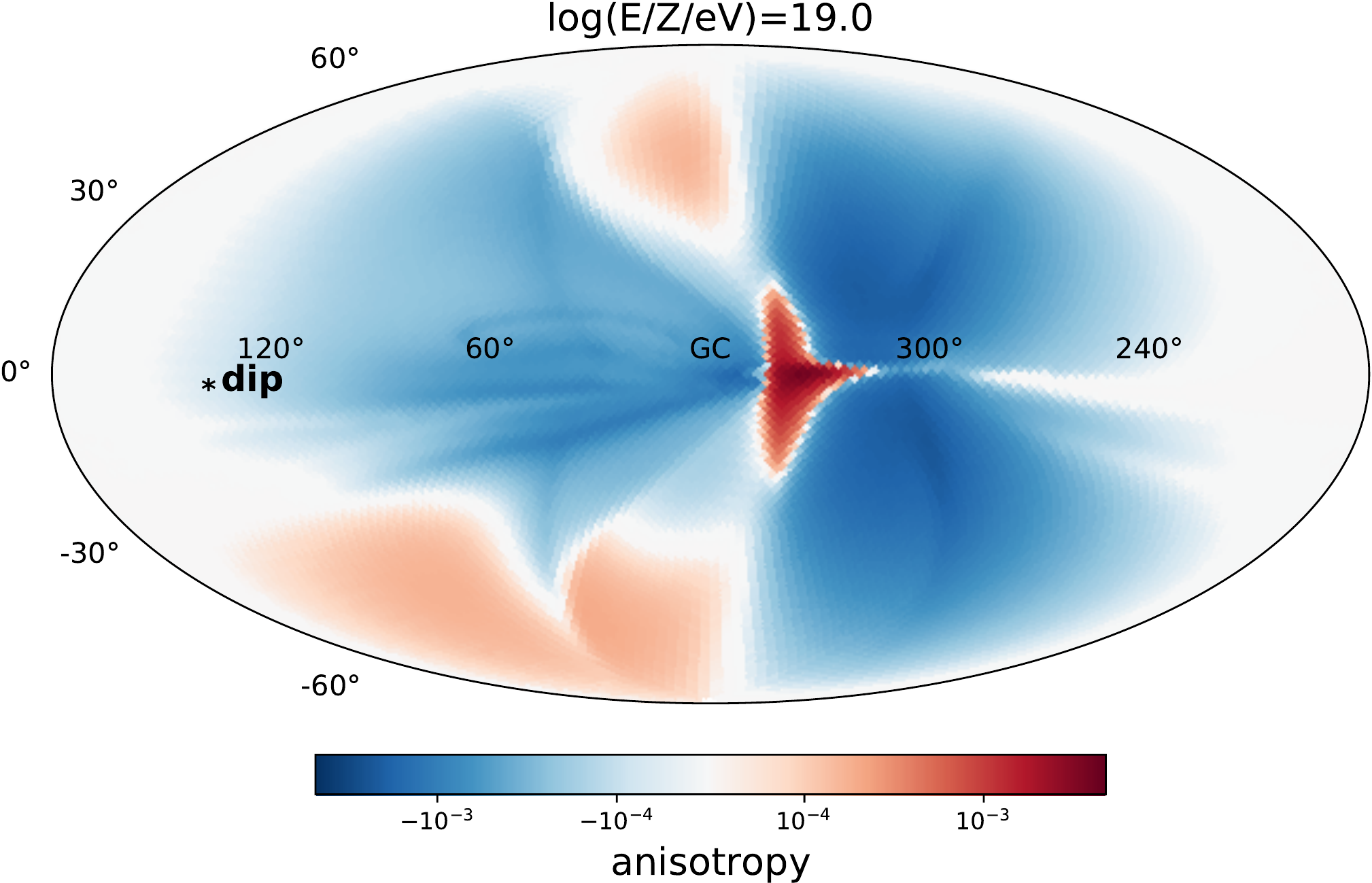} 
    \includegraphics[width=0.49\textwidth]{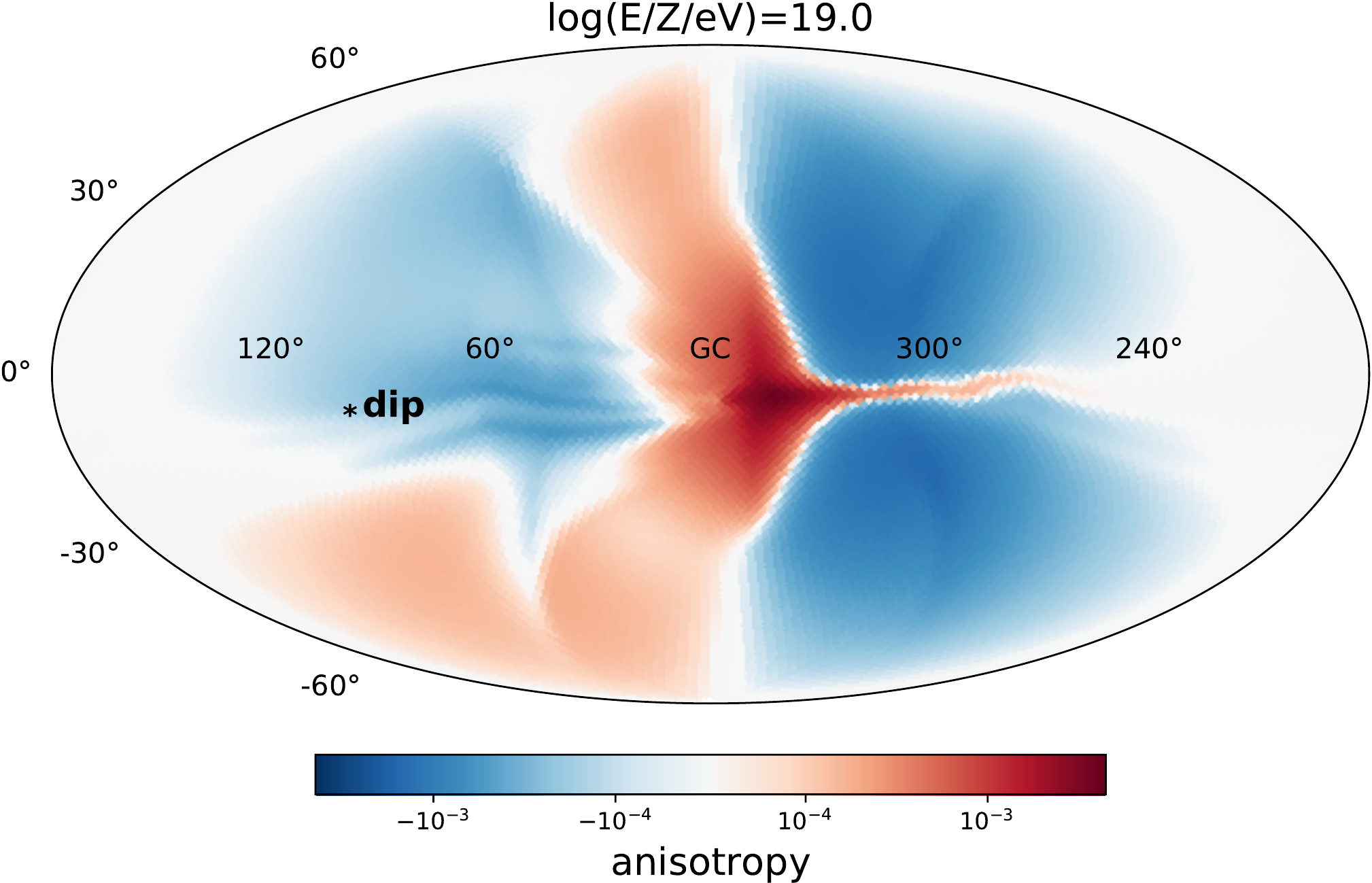} 
    
    \includegraphics[width=0.49\textwidth]{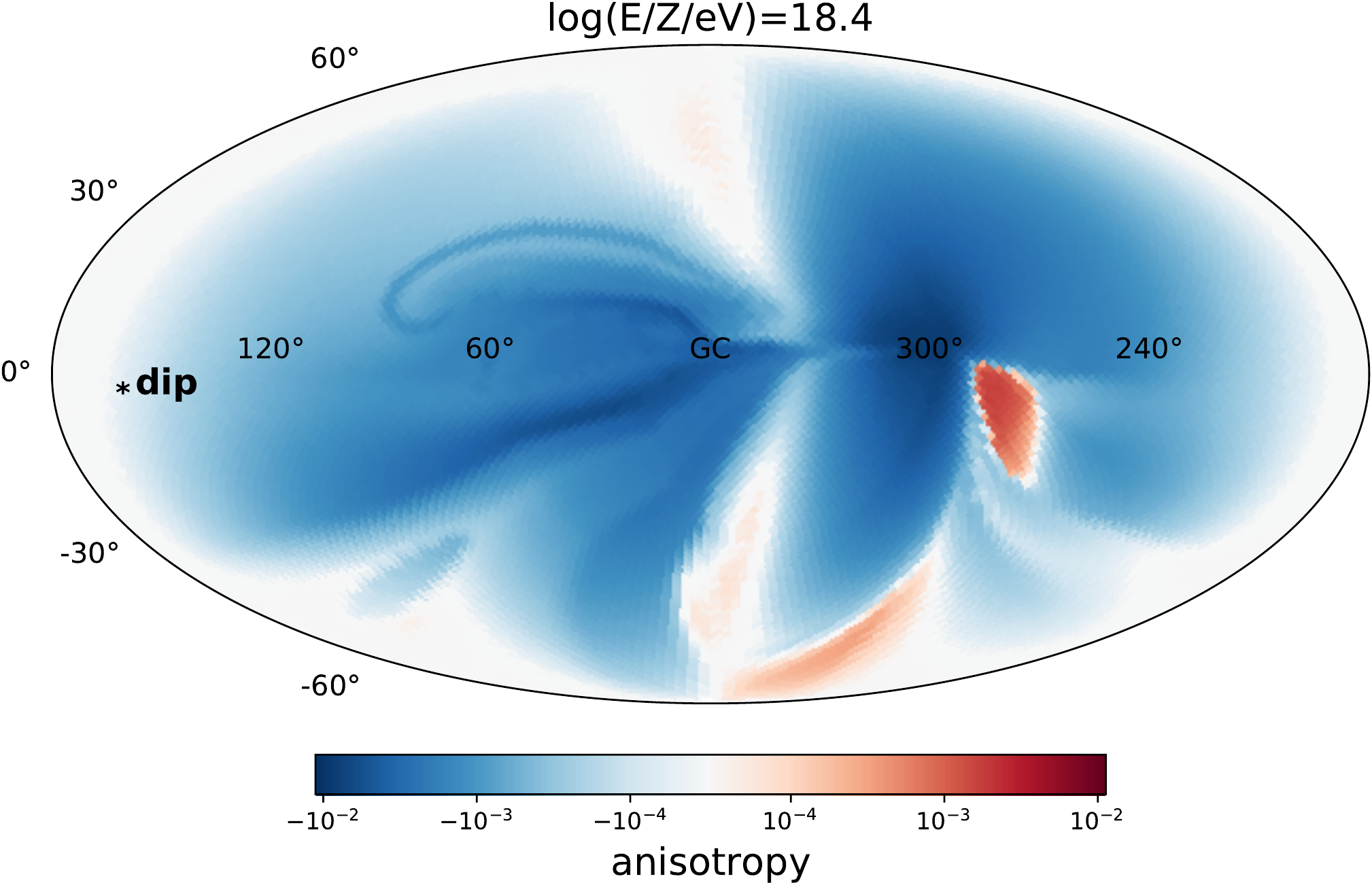} 
    \includegraphics[width=0.49\textwidth]{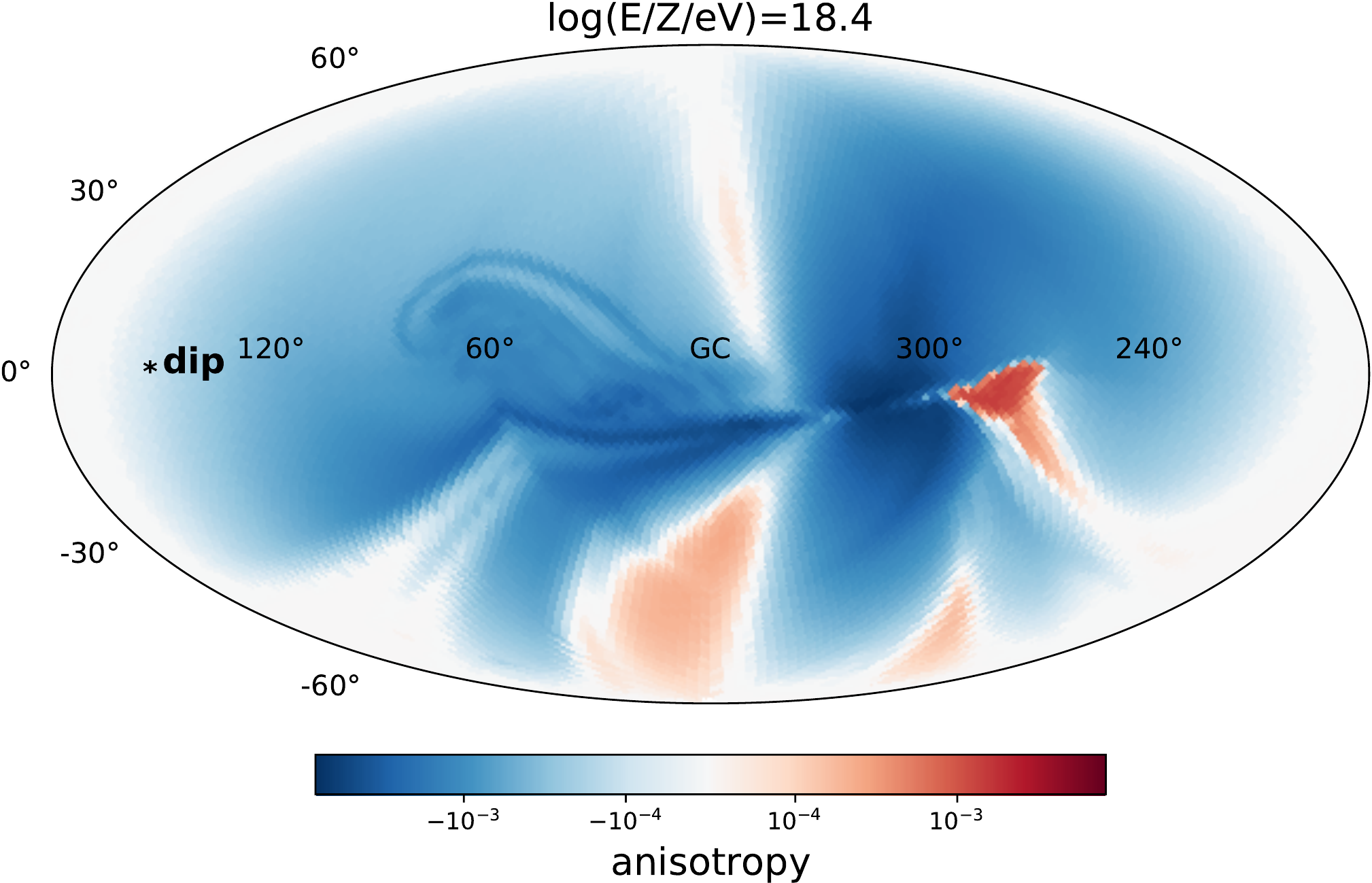} 

    \includegraphics[width=0.49\textwidth]{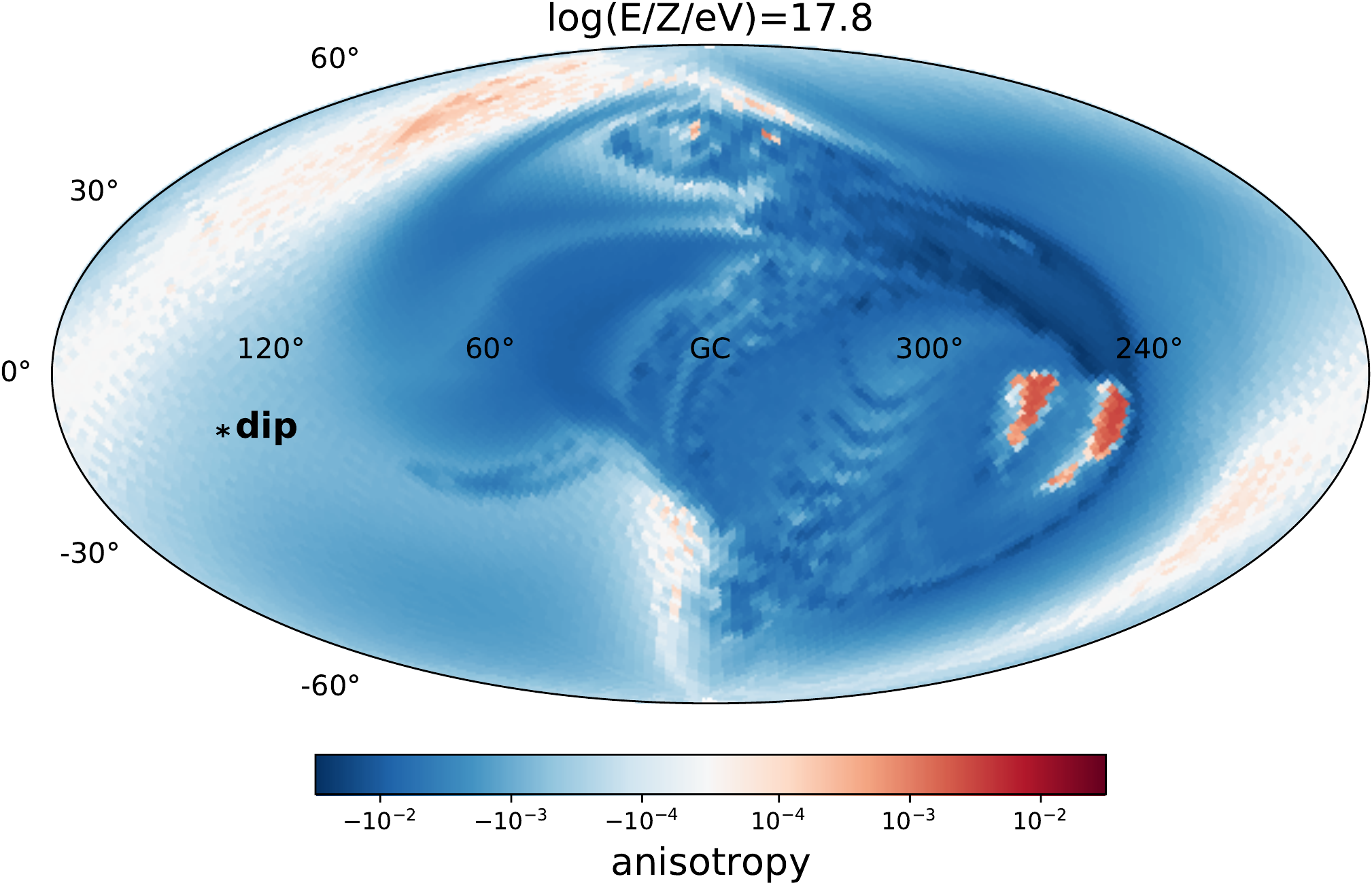} 
    \includegraphics[width=0.49\textwidth]{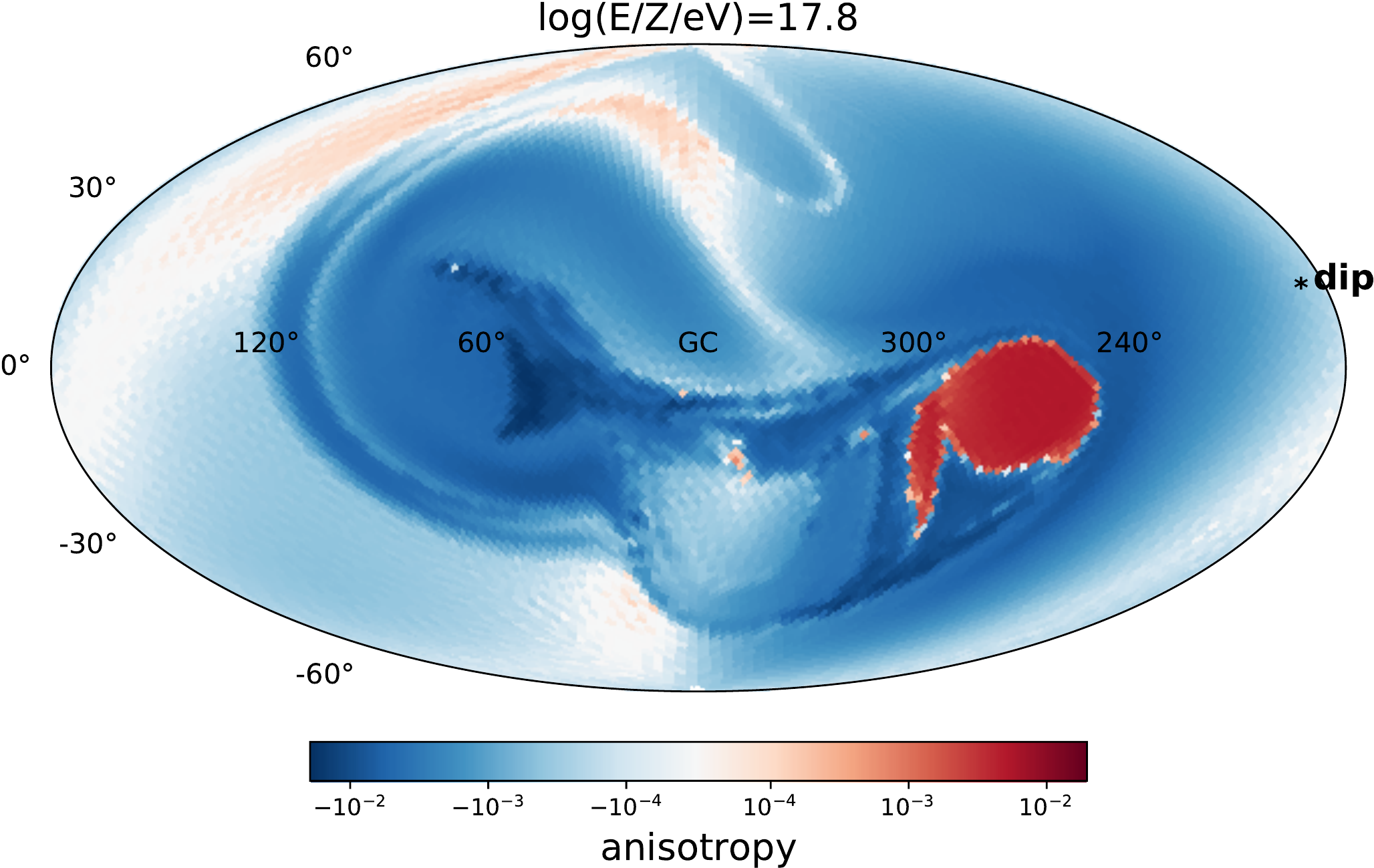} 

    \includegraphics[width=0.49\textwidth]{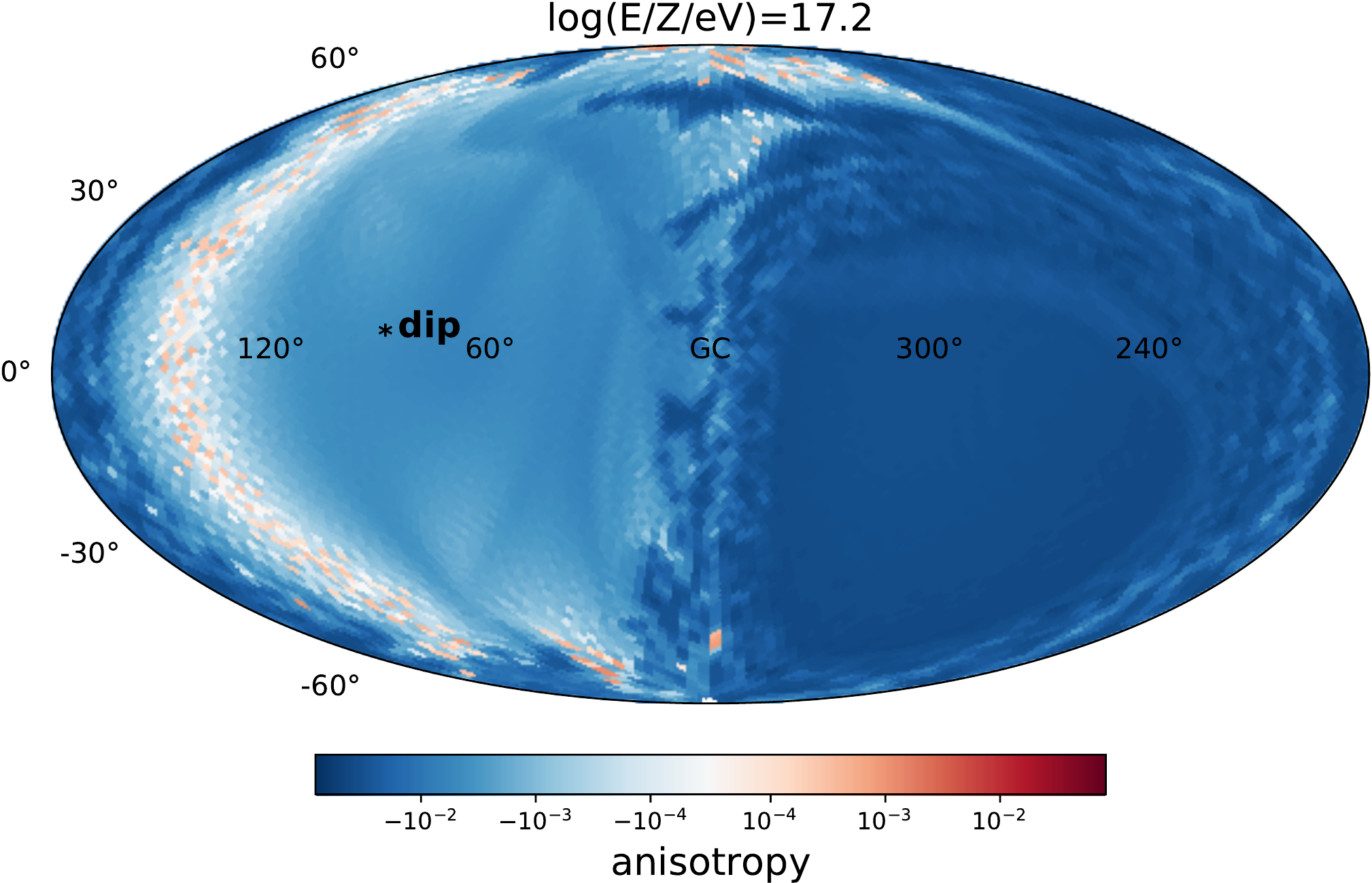} 
    \includegraphics[width=0.49\textwidth]{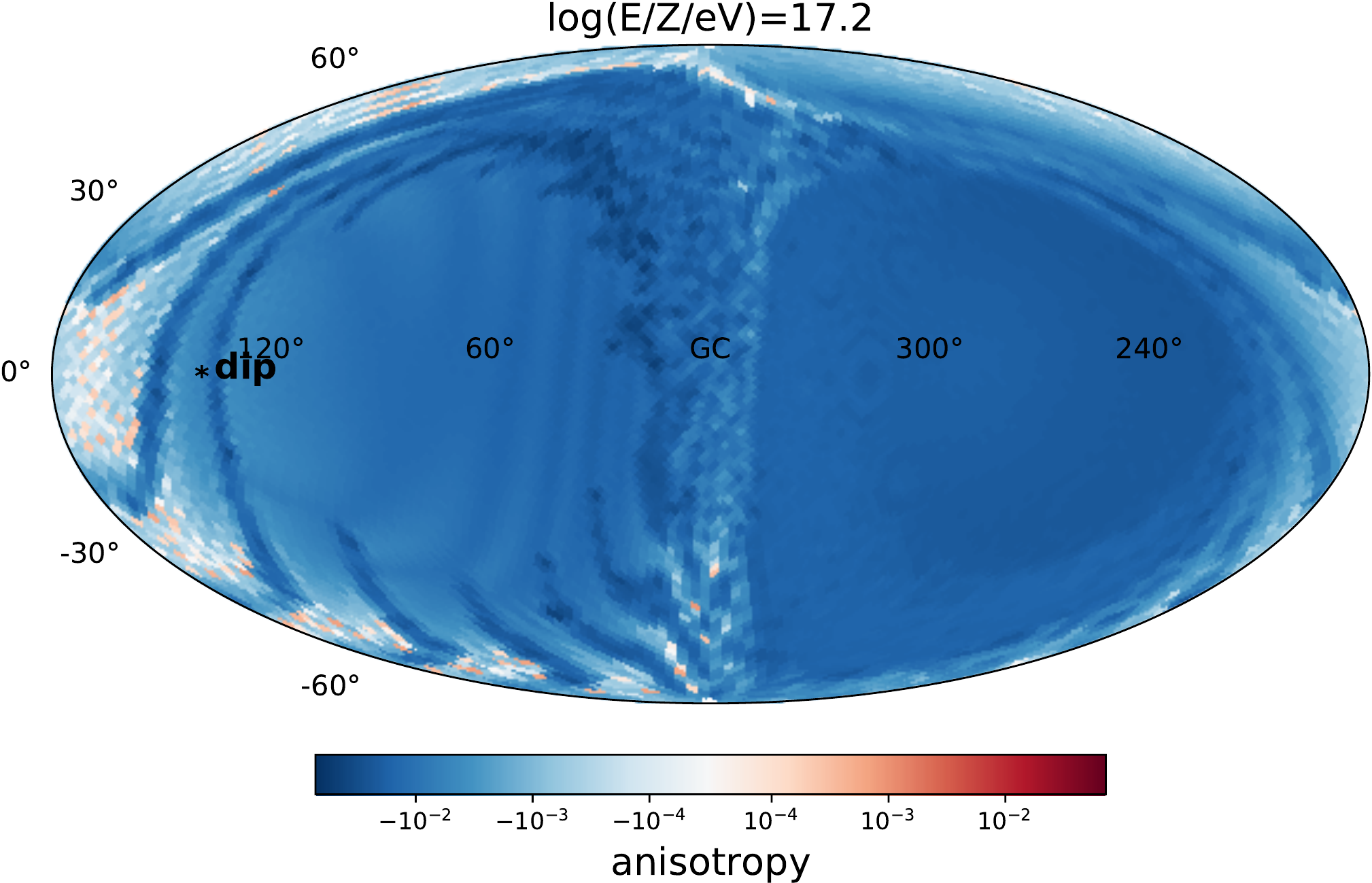} 

    \caption{Maps in Galactic coordinates of the anisotropies at Earth resulting from the acceleration of CRs due to the electric force, considering only the regular Galactic magnetic field and for different rigidities $E/Z$. The left column corresponds to the magnetic field model JF12, while the right column corresponds to JF+Planck. The direction of the dipolar component is indicated by an asterisk and labeled as `dip'.}
    \label{mapsBreg}
\end{figure}

Let us first discuss  the anisotropies resulting from the acceleration of the CRs by the electric force described by the last term in Eq.\,(\ref{phiearth}), that can be computed by integrating Eq.\,(\ref{phalo}) along the different particle trajectories,  considering that the original extragalactic flux is isotropic. We show in Fig.~\ref{mapsBreg} the anisotropy maps obtained by back tracking a regular grid of directions at Earth up to the outskirts of the Galaxy (taken as a sphere with galactocentric radius of 20~kpc) for different values of the CR rigidity at arrival,  $E/Z$, and considering a spectral index $\gamma = 3.3$, as is measured below the ankle \cite{Augerspec}. In these plots only the regular component of the magnetic field is included. The plots on the left  correspond to the results obtained using the JF12  magnetic field model, while those on the right  correspond to the results obtained with the modified field JF+Planck. Red regions in the maps correspond to the arrival directions of trajectories along which cosmic rays have been accelerated while traversing the Galactic magnetic field, and thus a larger flux should reach the Earth with the quoted energy as a consequence of the larger abundance of the particles which had lower energies outside the Galaxy. Blue regions instead correspond to directions along which particles have been decelerated, and thus a lower flux should reach the Earth at that energy given the lower abundance of  particles with  higher energies  outside the Galaxy. We can see that the shape of the anisotropies depends on the magnetic field model considered while the overall amplitudes of the resulting anisotropies are comparable in both models. 

At the highest value of $E/Z$ considered of $10^{19}$\,eV, for which the trajectories are less deflected by the magnetic field, we can identify more easily the main magnetic field component that is responsible for the anisotropy pattern observed. Since both the relative velocity
$\Delta\vec V \equiv \vec V - \vec V_\odot$ as well as the disk and toroidal magnetic field components are parallel to the Galactic plane, these components give rise to an electric field which is orthogonal to the Galactic plane. Thus, particles reaching the Earth along directions close to the Galactic plane are not significantly accelerated by these components, and the X-field component is the one giving the dominant contribution, having the effect of accelerating the particles coming through the inner Galactic directions, both above and below the Galactic plane. At higher Galactic latitudes, both towards the North and  the South of the Galactic plane, it is the toroidal halo component the one giving the largest contribution, involving a sign reversal of the effect from positive to negative longitudes. On the other hand, in this case there is   no sign reversal from the North with respect to the South because  the toroidal field component has opposite directions in the two hemispheres. For lower rigidities the CR trajectories   experience larger deflections, and hence the interpretation of these effects becomes  less direct.

\begin{figure}[h]
    \centering

    \includegraphics[width=0.49\textwidth]{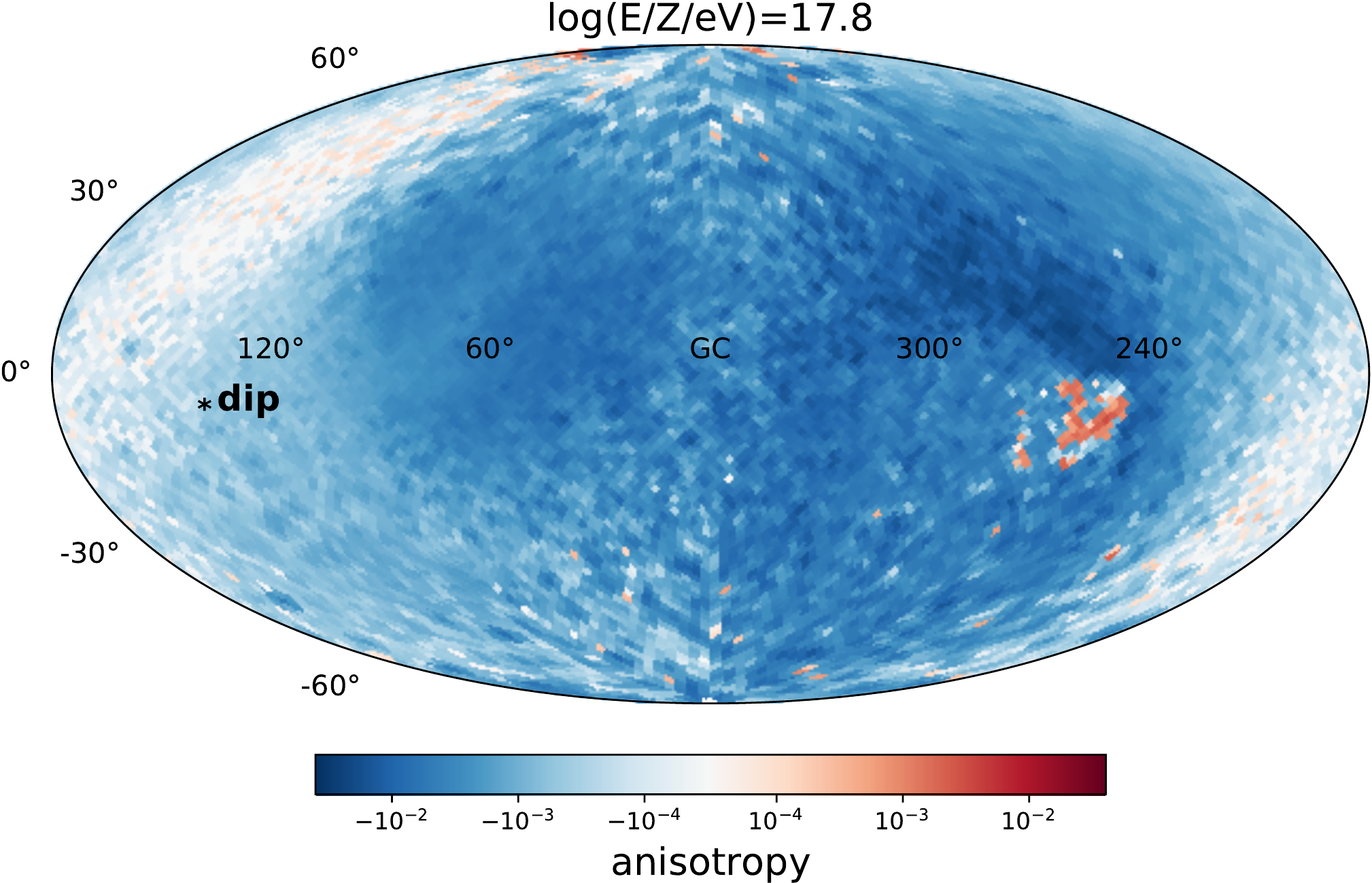} 
    \includegraphics[width=0.49\textwidth]{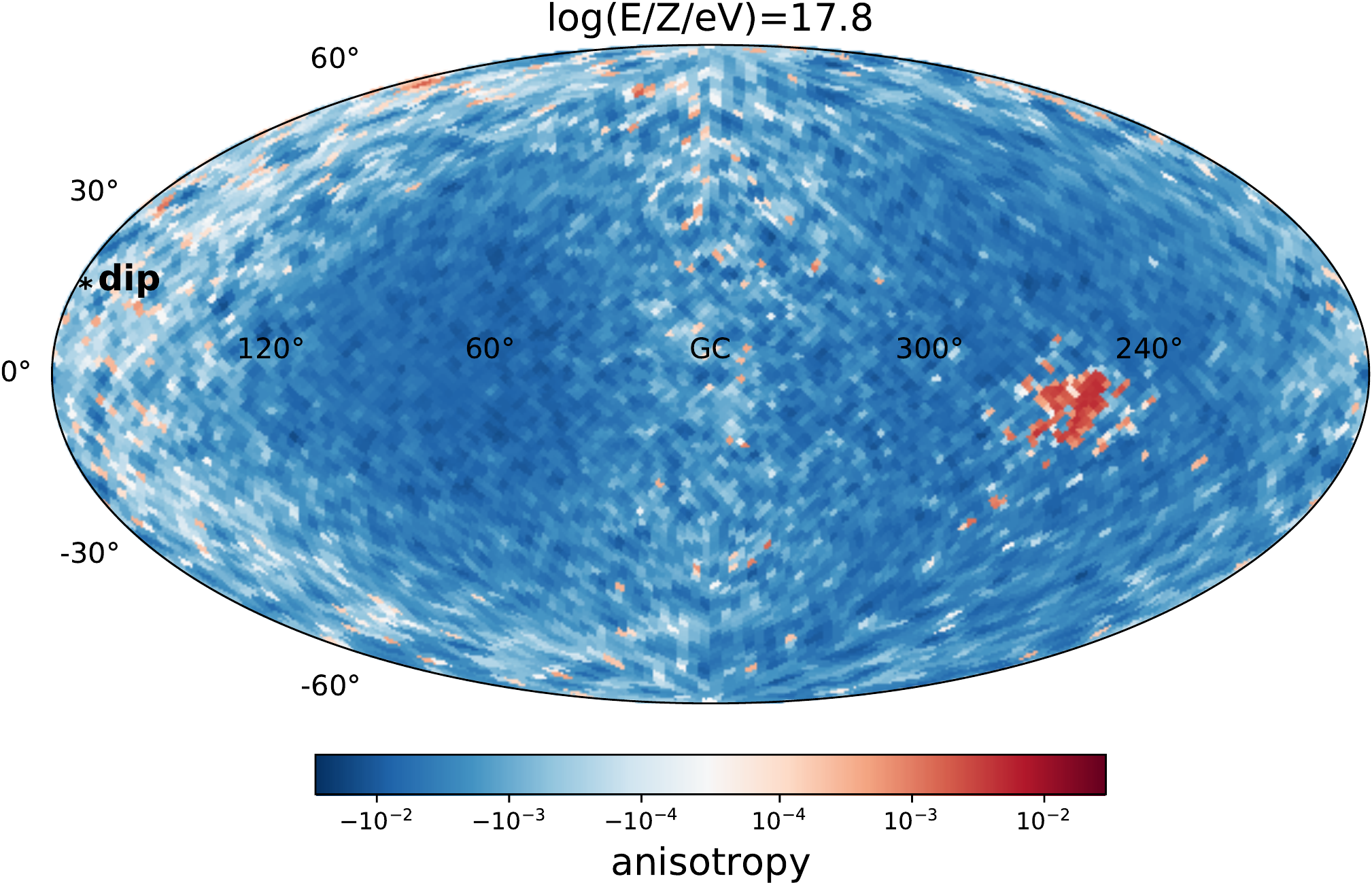} 
    \includegraphics[width=0.49\textwidth]{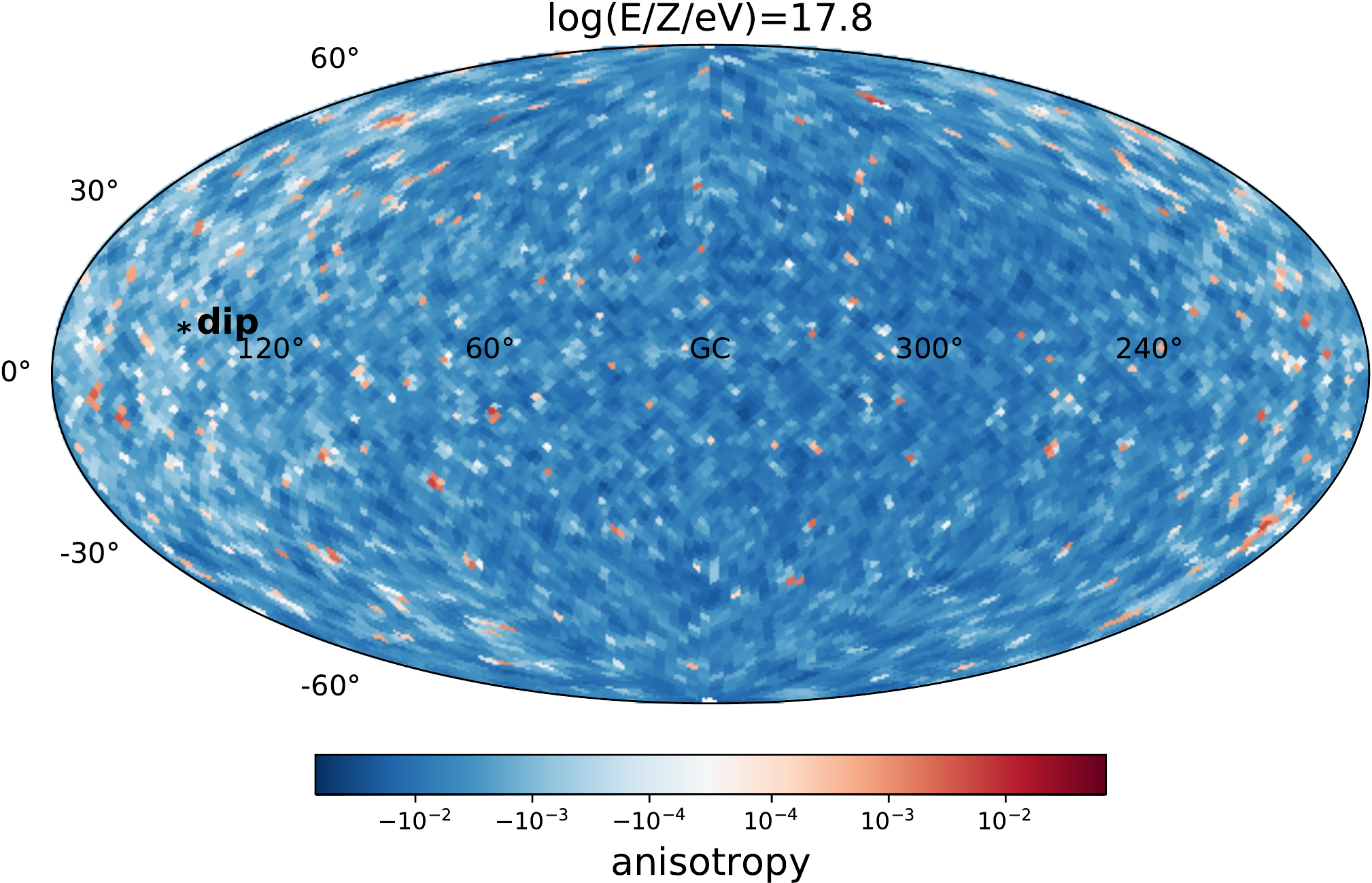} 
    \includegraphics[width=0.49\textwidth]{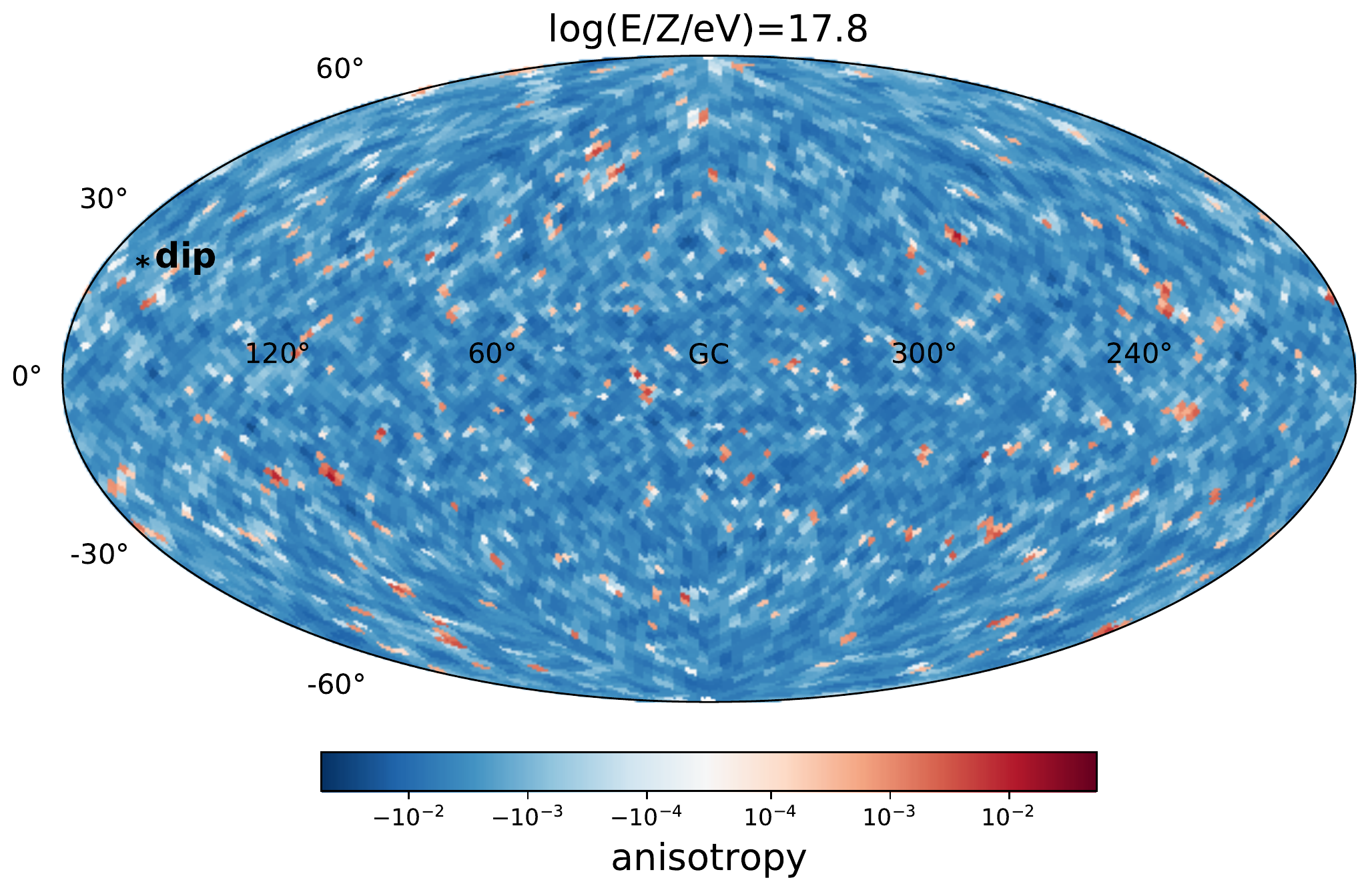} 
   
    \caption{Maps in Galactic coordinates of the anisotropy  at Earth from the acceleration of CRs due to the electric force, considering the regular and random striated magnetic field components (top row) and including also the random isotropic component (bottom row), for a rigidity $E/Z=10^{17.8}$\,eV. The left column corresponds to the magnetic field model JF12 while the right column corresponds to JF+Planck.}
    \label{mapsBran}
\end{figure}

Including the random striated and random isotropic magnetic field components has  in both cases the effect of smoothing the features that were present in the  maps including only  the regular field. 
We show in Fig.~\ref{mapsBran} the effect of the random field components for both the JF12 and JF+Planck, for a rigidity of $10^{17.8}$~eV. As in Fig.~\ref{mapsBreg},  the left plots correspond to the JF12 model and the right ones to the JF+Planck model, while the top row shows the results including the regular and random striated components, and the bottom row the complete field including also the isotropic random component. 
The details of the maps actually depend on the particular realization of the random field deflections that were considered, and the differences between different realizations would become more significant when considering smaller rigidities, for which  the deflections  are larger.

\begin{figure}[h]
    \centering
    \includegraphics[width=0.49\textwidth]{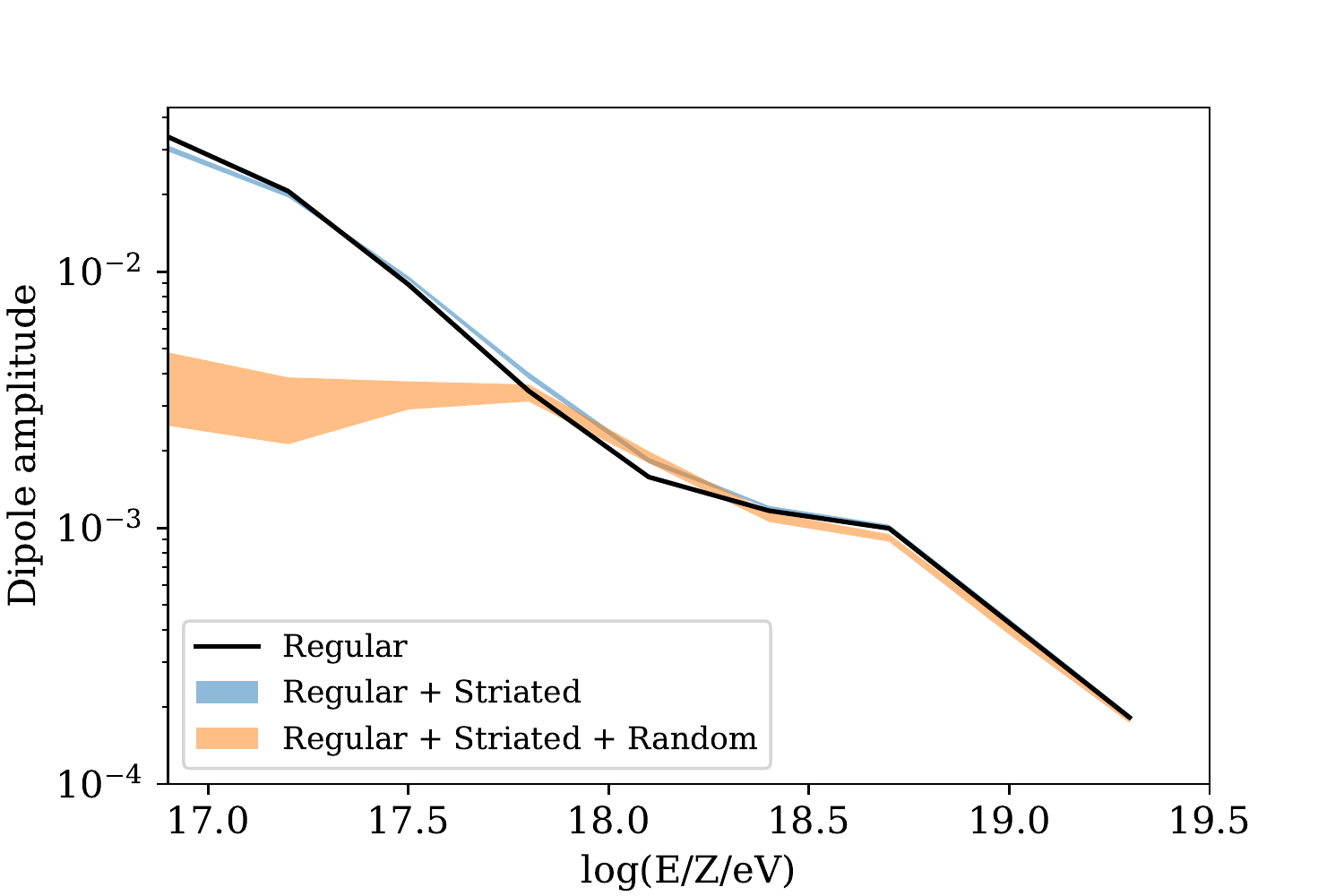} 
    \includegraphics[width=0.49\textwidth]{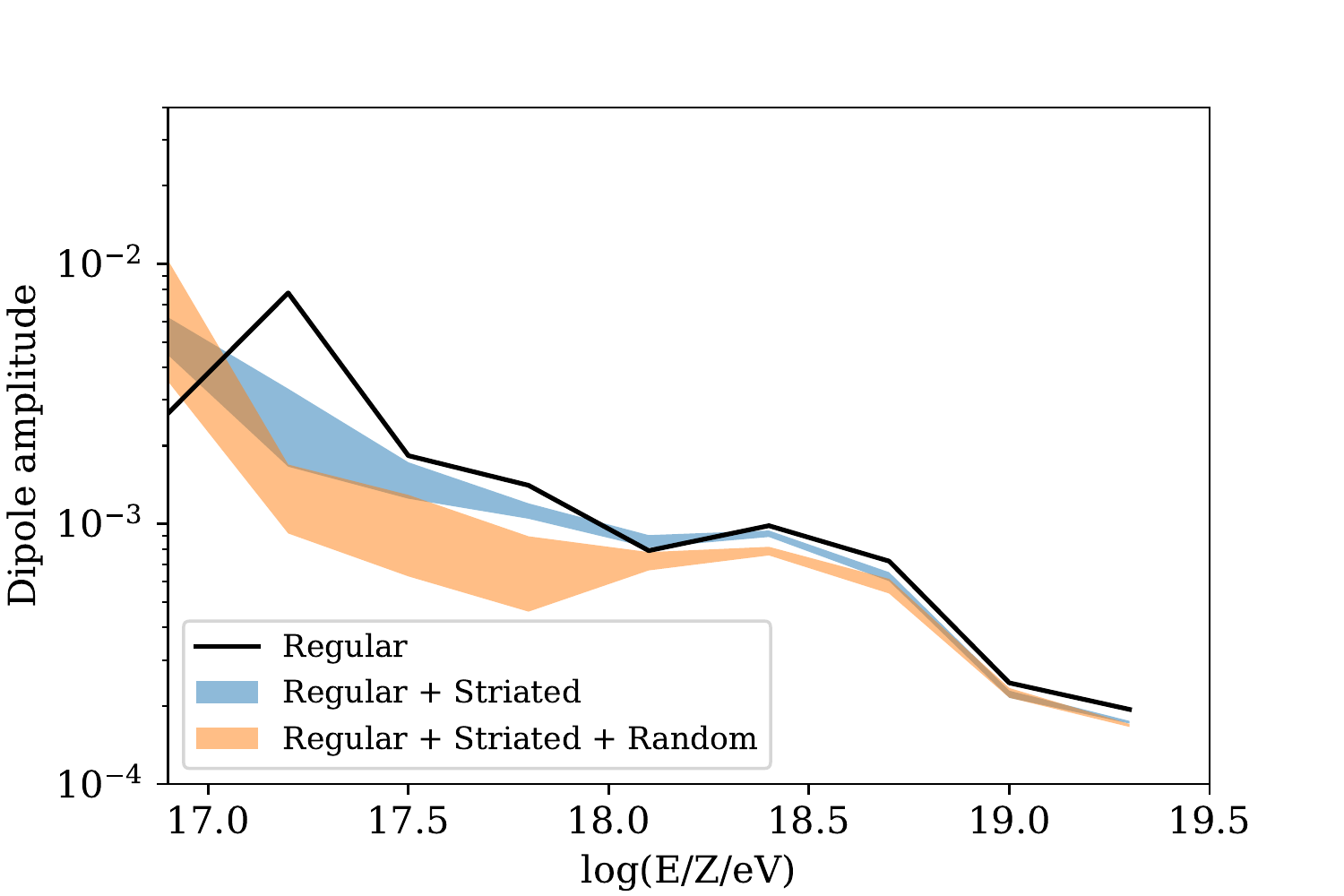} 
    \caption{Dipole amplitude as a function of the logarithm of the rigidity $E/Z$. Left panel corresponds to the JF12 magnetic field model while the right panel corresponds to JF+Planck.}
    \label{ampvsE}
\end{figure}

Of particular interest is the contribution due to the acceleration by the electric force to the dipolar component of the flux distribution at  Earth. The  direction of the associated dipole is shown with an asterisk in the examples of Figs.~\ref{mapsBreg} and \ref{mapsBran} (and is labeled as `dip').  In Fig.~\ref{ampvsE} we show the amplitude of the dipolar component as a function of the rigidity $E/Z$, obtained when considering only the regular field, the regular and the random striated components and the complete field including also the random isotropic component. For the cases with random fields the colored bands correspond to the dispersion around the mean values obtained from ten different simulations for each energy. As already mentioned, for both models the dipolar amplitude due to the CR acceleration tends to increase for decreasing rigidity, while on the other hand the inclusion of the random field leads, due to the smearing induced, to a reduction of the dipolar component. The amplitude resulting from the regular magnetic field model is slightly smaller for the JF+Planck model, mostly as a consequence  of the reduced magnitude of the halo field. The effect of the striated field on the dipolar component is more notorious in the JF+Planck model due to the larger associated value of the factor $\beta$, and it reduces the amplitude of the dipolar component at low energies.

\begin{figure}[h]
    \centering
    \includegraphics[width=0.49\textwidth]{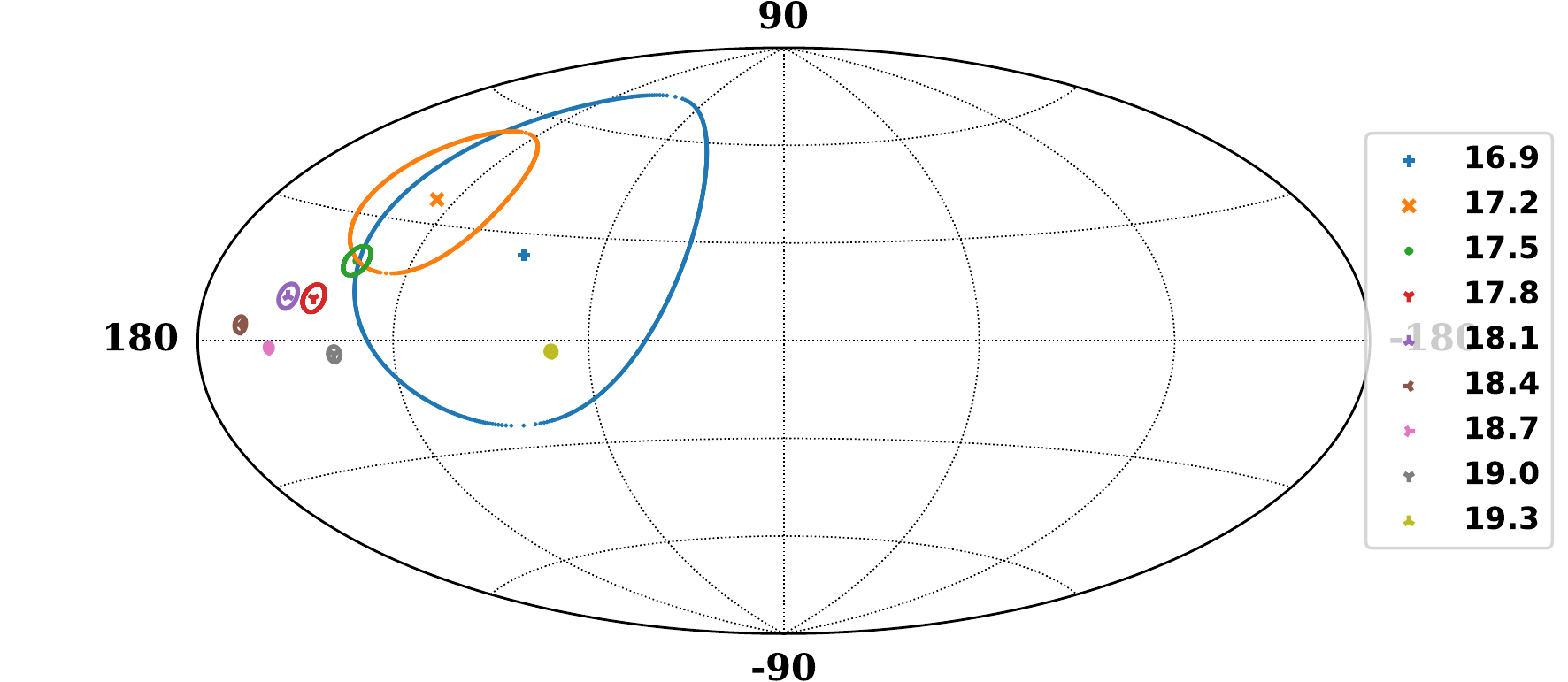} 
    \includegraphics[width=0.49\textwidth]{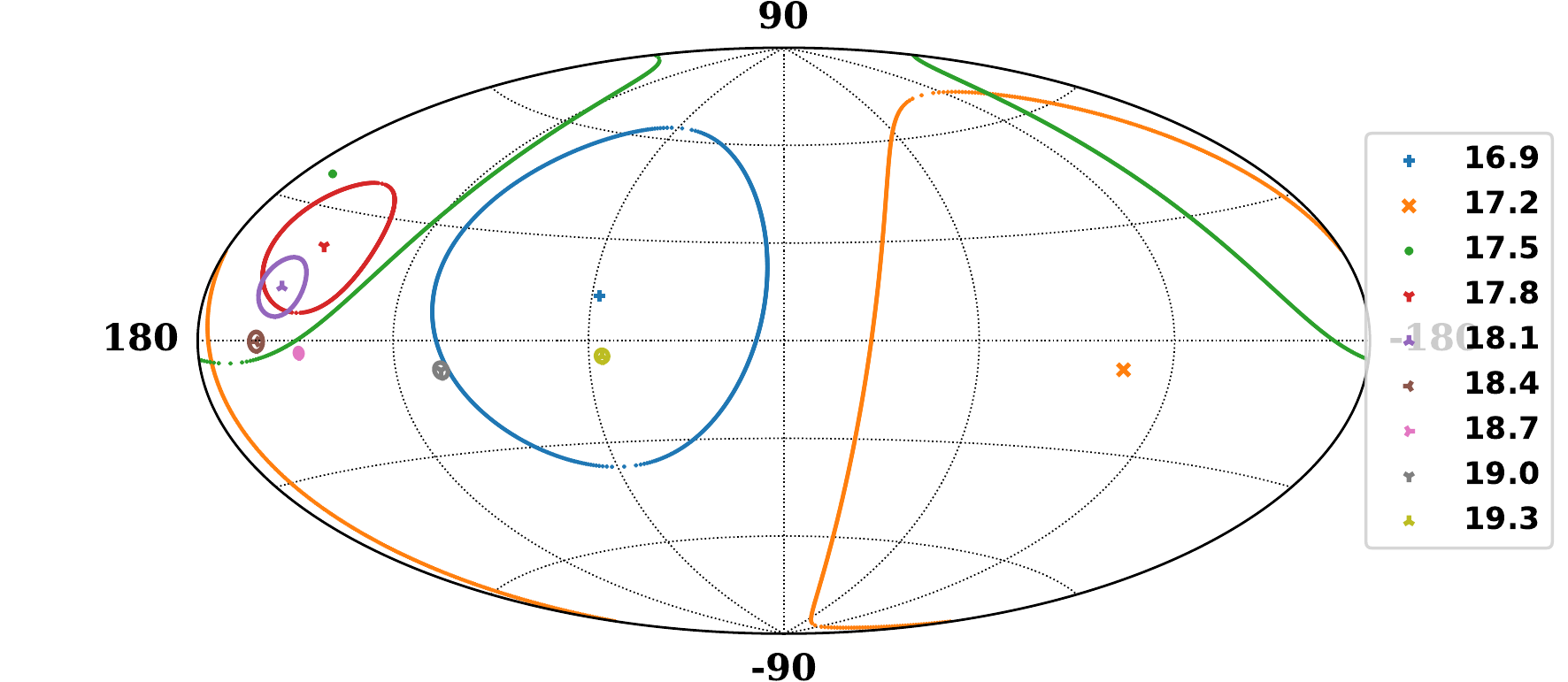}
    \caption{Maps in Galactic coordinates showing the average dipole direction for different values of log($E/Z/$eV) when the complete  magnetic field (regular + striated + random isotropic components) is considered.  The circular contours indicate the median of the distribution of the dipole directions. Left panel corresponds to the JF12 magnetic field model, right panel corresponds to JF+Planck.}
    \label{dirvsE}
\end{figure}

Figure~\ref{dirvsE} shows with dots the mean dipole direction obtained in several simulations with different realizations of the random deflections, and also shows the circular sky region around it within which the dipole points in half of the simulations, for different rigidities and considering the complete (regular, striated and random isotropic) field. At the highest energies, for which the dominant effect is that of the regular field, a small dispersion in the directions (as well as in the amplitudes shown in Fig.~\ref{ampvsE}) appears. The trend is similar for the JF12 model, displayed in the left panel, and for the JF+Planck one, displayed in the right panel. As the energy decreases and the effect of the random magnetic field becomes relatively more important, the dispersion in the dipolar directions increases, being this more notorious for the JF+Planck model due to the larger amplitude of the halo random isotropic component in this case. In particular, at the lowest energy displayed,  the random isotropic component is dominant for this model, resulting in an almost random distribution of the dipolar directions in the different simulations.

From the plots in Figs.~\ref{mapsBreg} and \ref{mapsBran} it is apparent that the anisotropies have not just a dipolar component, but that higher order multipole moments are also significant. 
By expanding the flux reaching the Earth in terms of spherical harmonics with coefficients $a_{\ell m}$, such that\footnote{We use the real spherical harmonic basis, so that the $a_{\ell m}$
are real.}
\begin{equation}
\Phi(\hat u)=\sum^{\infty}_{l=0}\ \sum^{\ell}_{m=-\ell}\
a_{\ell m}\ Y_{\ell m}(\hat u),
\end{equation}
we can quantify the amplitude of the different multipoles
through the angular power spectrum $C_\ell$, which is obtained by  averaging  the $a_{\ell m}^2$ coefficients 
over the $2\ell +1$ possible values of $m$
\begin{equation}
C_\ell = \frac{1}{2\ell+1}\ \sum_{m=-\ell}^{\ell}\ a_{\ell m}^2.
\end{equation}

\begin{figure}[h]
    \centering
    \includegraphics[width=0.49\textwidth]{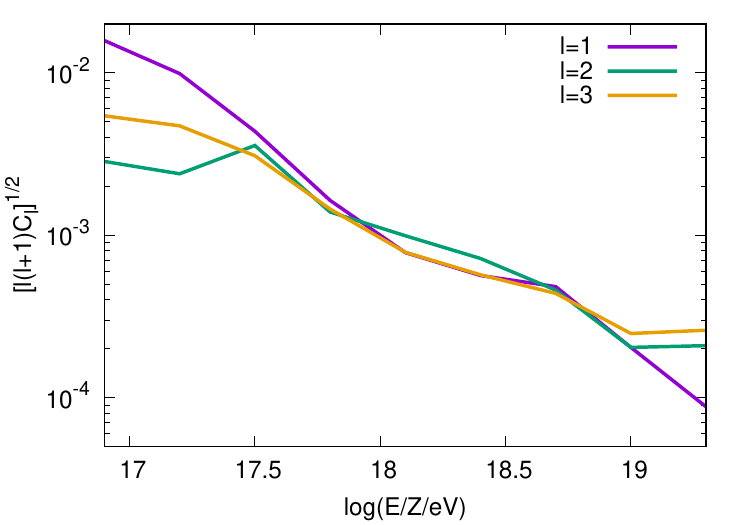} 
    \includegraphics[width=0.49\textwidth]{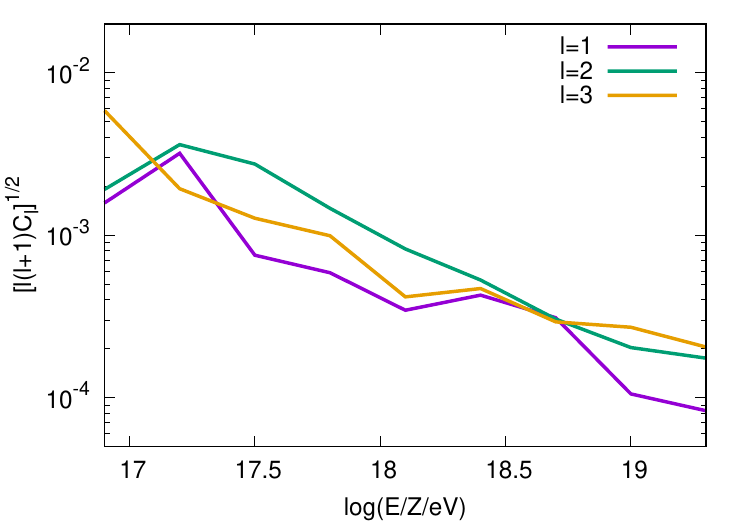} 
    \caption{Normalized multipole amplitude as a function of the rigidity for the three lowest $\ell$ values of the anisotropies resulting from the acceleration due to the electric force. The left panel corresponds to the regular component of the  JF12 magnetic field model, while the right panel corresponds to JF+Planck.}
    \label{clEF}
\end{figure}

We show in Fig.~\ref{clEF} the normalized contribution $\sqrt{\ell(\ell+1) C_\ell}$ for the three lowest order terms, which are the dipole, the quadrupole and the octupole, as a function of the rigidity, when considering the regular magnetic field component of the JF12 and JF+Planck models. The amplitude of all of them are of  similar size and they decrease  with increasing energy, ranging from ${\rm few} \times 10^{-3}$ for rigidities close to $10^{17}$~eV down to $\sim 10^{-4}$ for rigidities above $10^{19}$~eV.

\subsection{Modification of the anisotropies outside the Galaxy due to Galactic magnetic field deflections}
\label{comptongetting}

In addition to the anisotropies due to the acceleration by the electric force just discussed, other sources of anisotropies of extragalactic CRs could be the Compton-Getting effect resulting from the motion of the observer with respect to the frame in which CRs are isotropic (second term in Eq.~(\ref{phiearth})) or eventually  some intrinsic anisotropies may already be present in the flux reaching the halo of the Galaxy. 

\begin{figure}[h]
    \centering
    \includegraphics[width=0.49\textwidth]{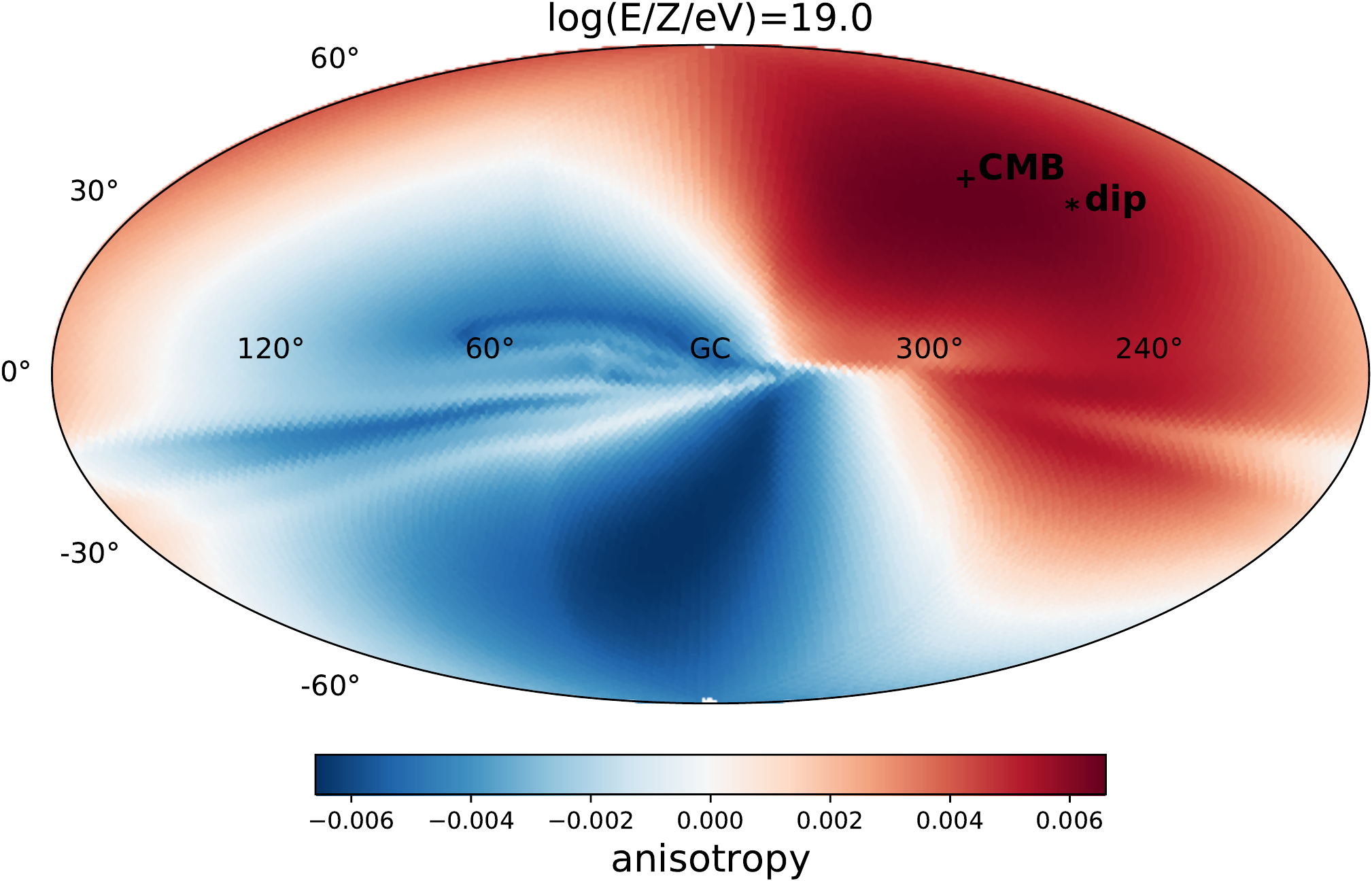} 
    \includegraphics[width=0.49\textwidth]{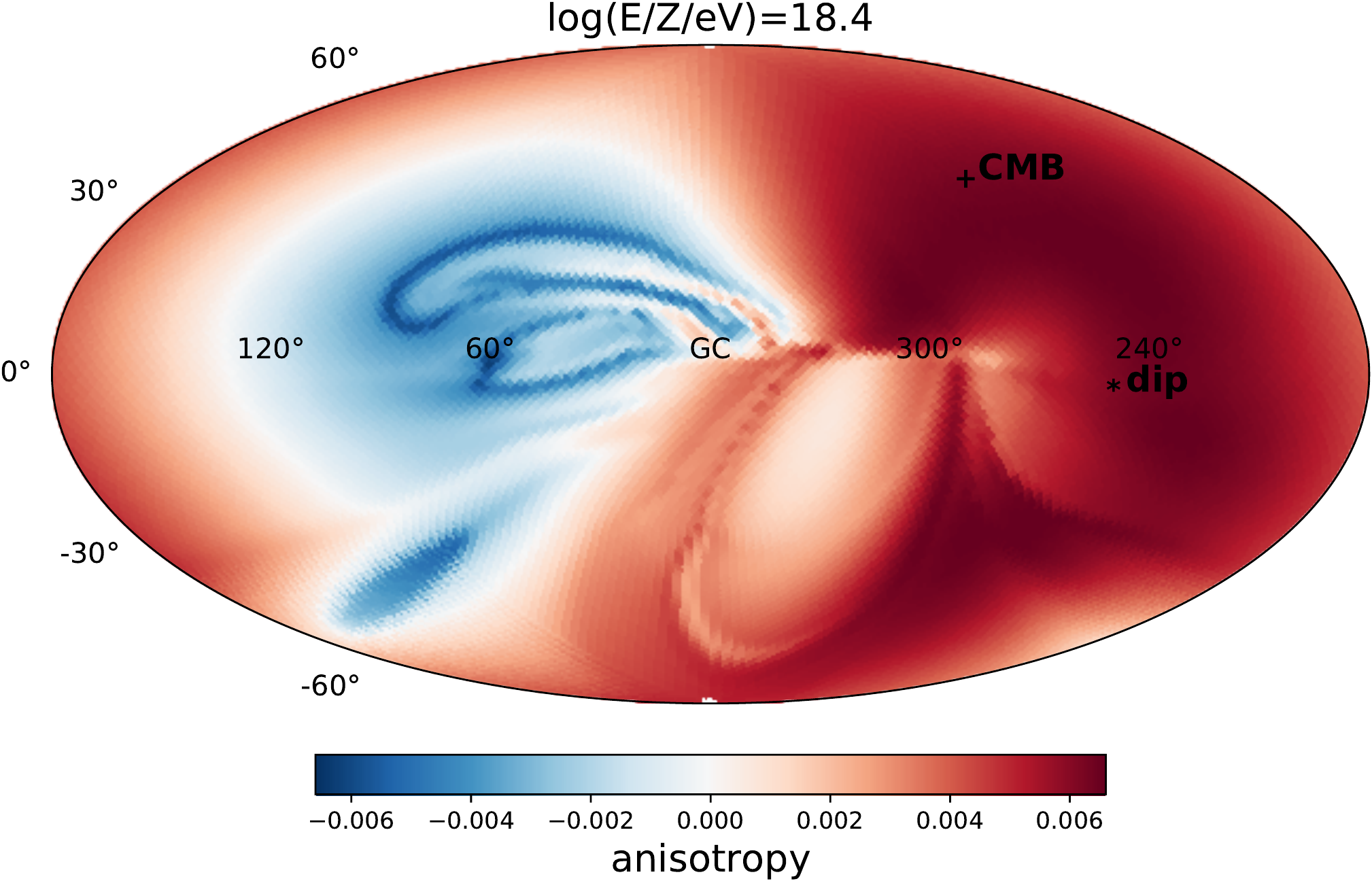} 
    
    \includegraphics[width=0.49\textwidth]{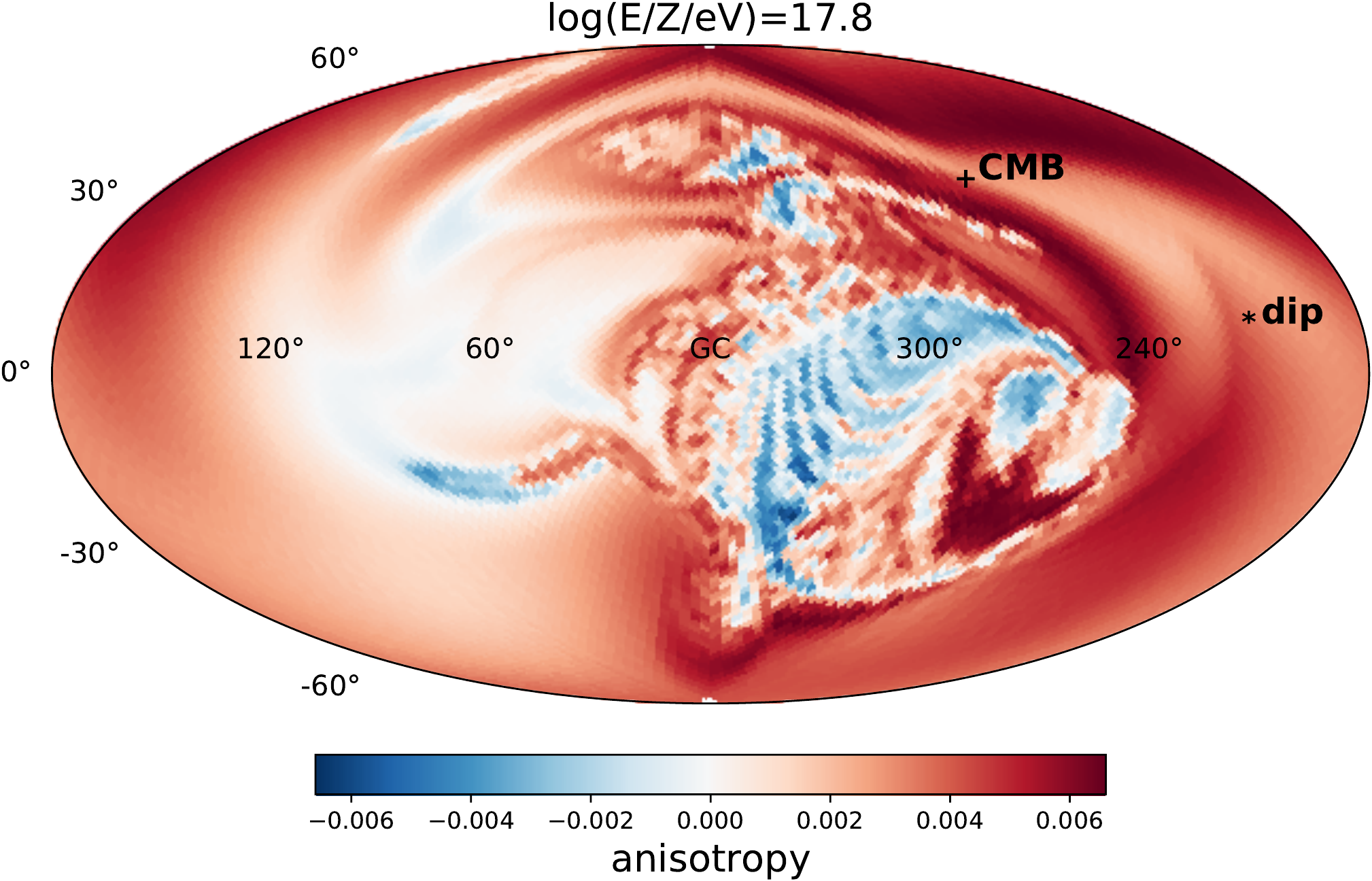} 
    \includegraphics[width=0.49\textwidth]{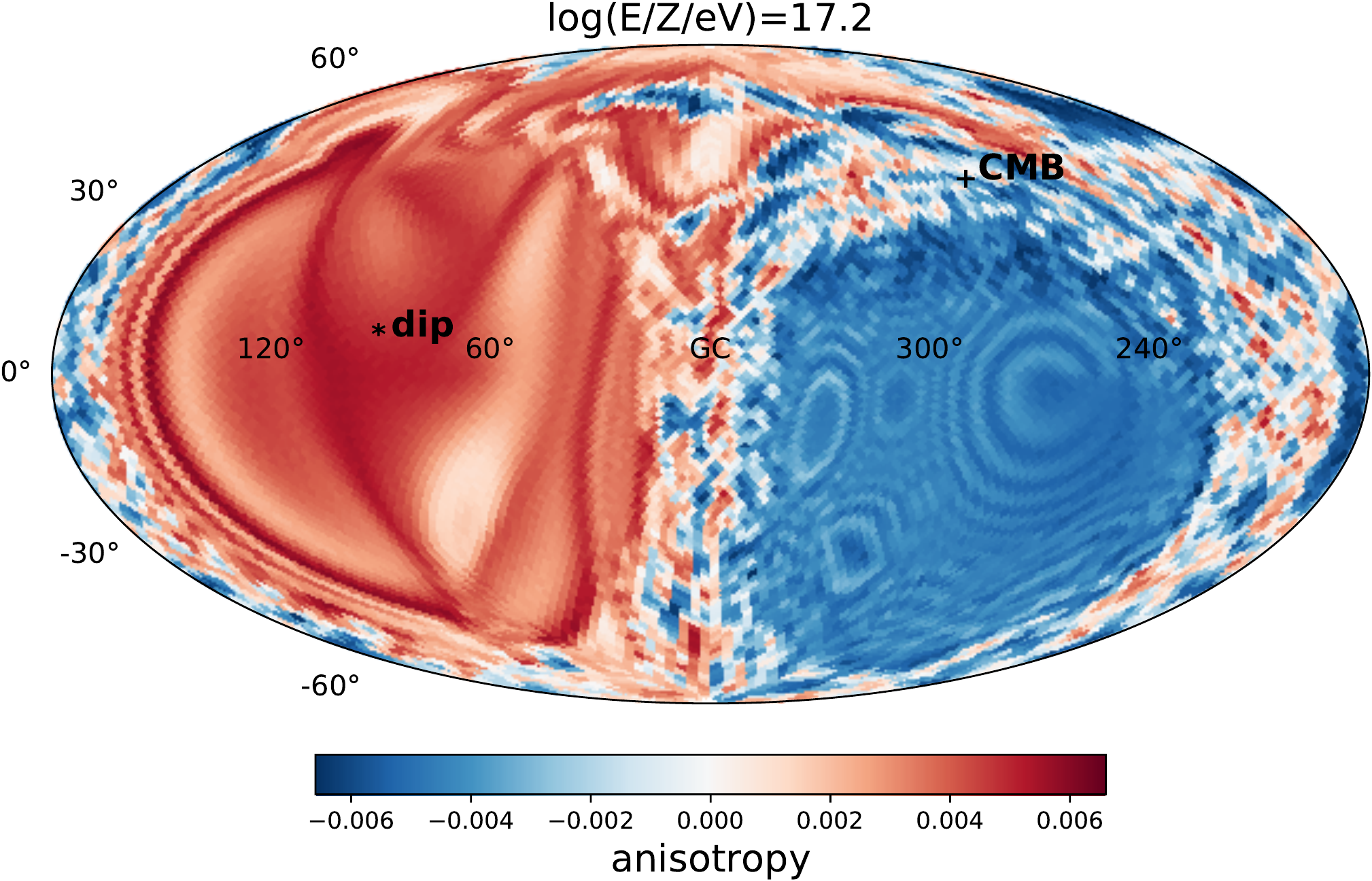} 
   
    \caption{Maps in Galactic coordinates of the anisotropies resulting from the Compton-Getting effect, for different values of the rigidity, assuming that the CRs are isotropic in the CMB rest frame and considering the regular Galactic magnetic field model of JF12. The direction towards the CMB dipole is indicated with a plus sign and that of the resulting CR dipole with a star.}
    \label{mapcgreg}
    \end{figure}

Setting aside possible intrinsic anisotropies and in the absence of the Galactic magnetic field (so that $\hat u_h = \hat u_0$), the Compton-Getting effect would give rise to a dipolar anisotropy pointing in the direction of the velocity of the observer with respect to the isotropic frame. For definiteness we will exemplify the distortion of this pattern resulting from the propagation of CRs through the Galactic magnetic field for the case in which the extragalactic CRs are isotropic in the rest frame of the CMB. Figure~\ref{mapcgreg} shows the resulting anisotropy maps at the Earth for different rigidities, obtained considering the effects of the regular magnetic field component of the JF12 model. At the highest rigidity displayed, the overall distribution is close to dipolar, pointing in  a direction close to the CMB dipole one and with an amplitude  $d \simeq 0.0064$.  For decreasing energies the amplitude of the dipolar component decreases to $d \simeq 0.0050$ for $E/Z=10^{18.4}$~eV and $d \simeq 0.0020$ for $E/Z=10^{17.8}$~eV, while the direction moves down towards the Galactic plane. At the smallest rigidity displayed, $E/Z=10^{17.2}$~eV, the amplitude grows again to $d \simeq 0.0056$ and the direction flips to the almost  opposite longitude.

\begin{figure}[h]
    \centering
    \includegraphics[width=0.49\textwidth]{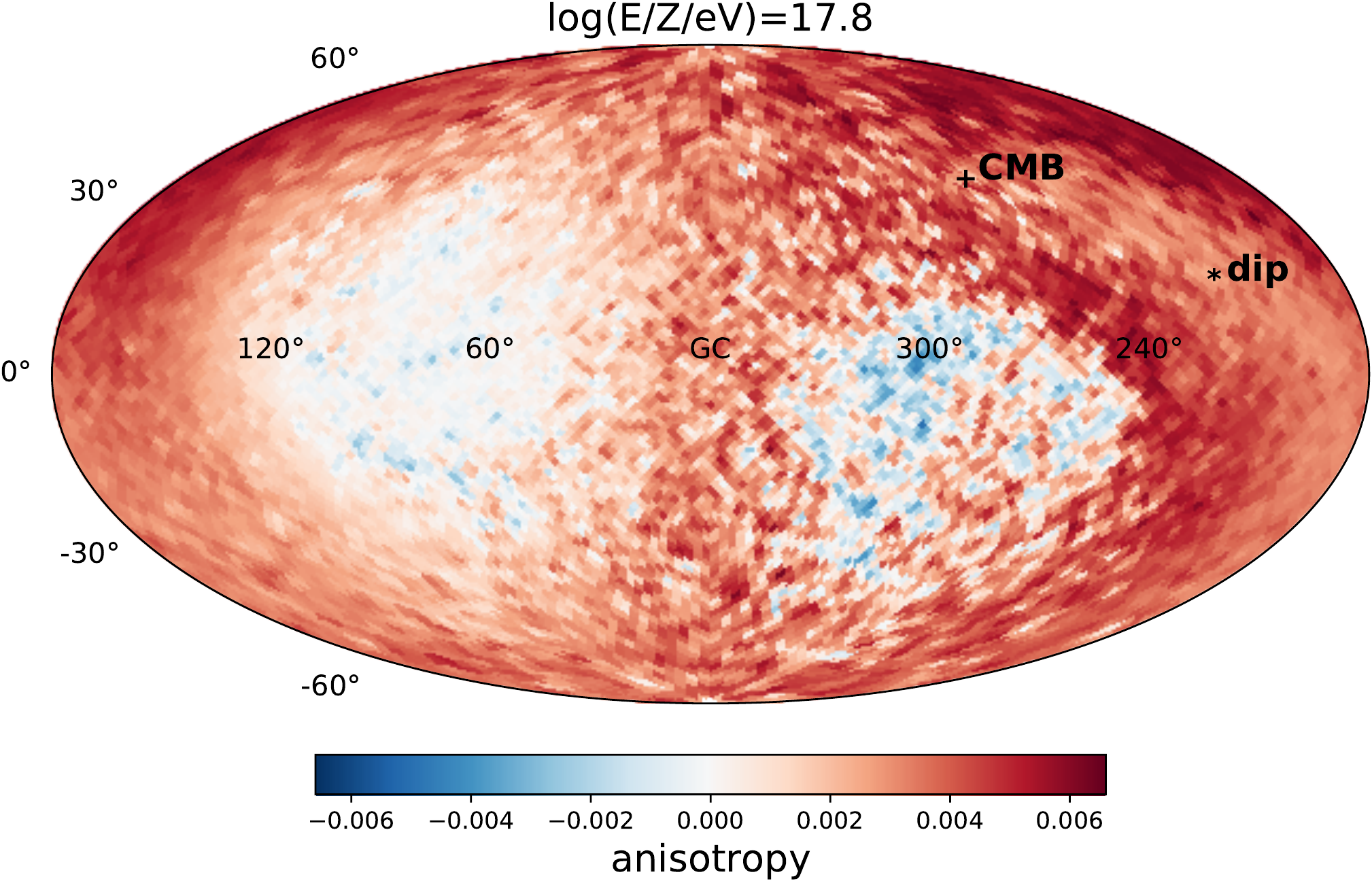} 
    \includegraphics[width=0.49\textwidth]{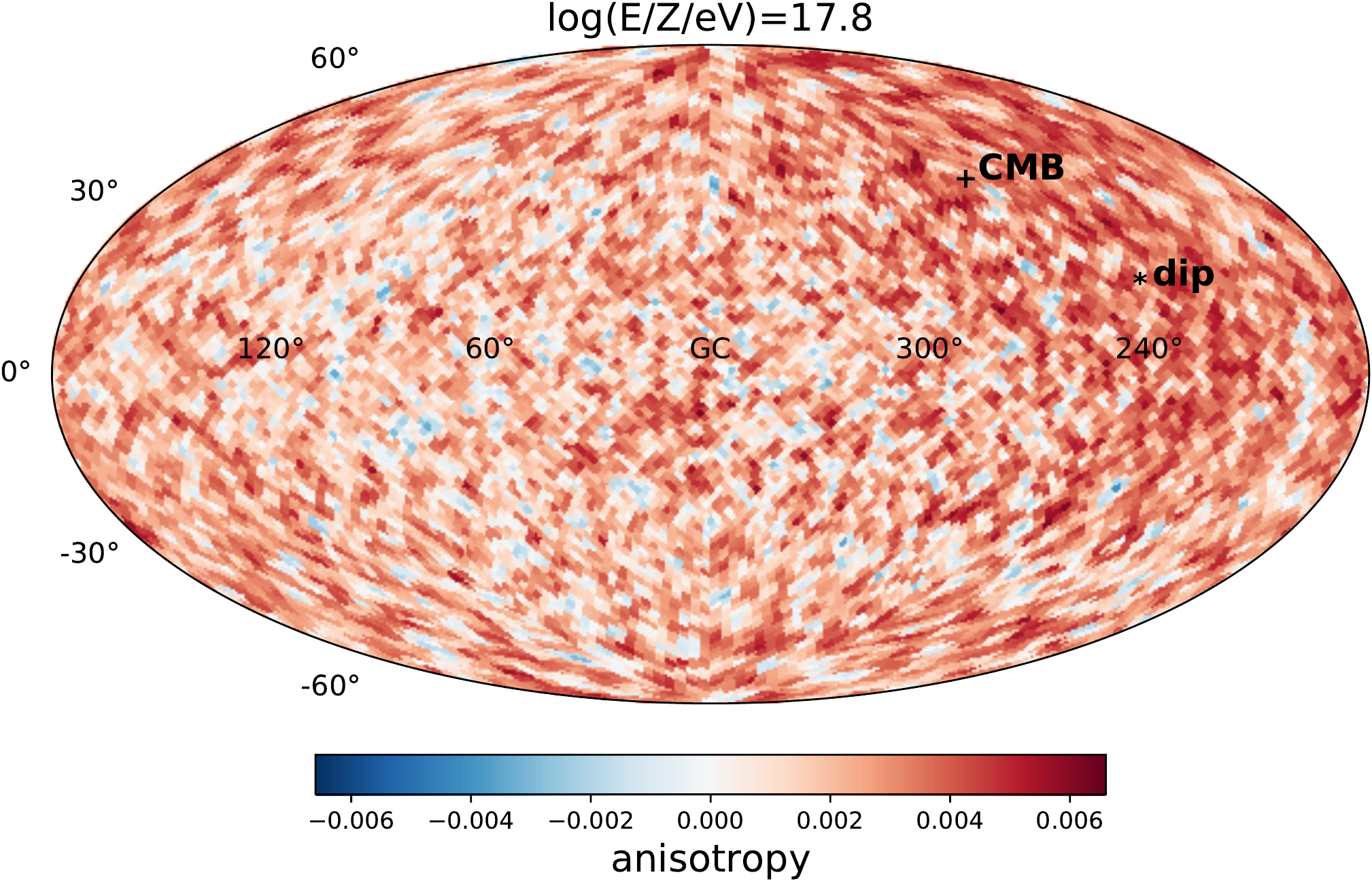} 
    
    \caption{Same as Fig.~\ref{mapcgreg} for a rigidity $E/Z = 10^{17.8}$~eV, considering the regular and striated magnetic field components (left panel) and the complete field including also the isotropic random component (right panel) of the JF12 model.}
    \label{mapcgran}
    \end{figure}

     The random striated and random isotropic components have  the effect of smoothing  down the anisotropies and of reducing the dipolar amplitude. We show in Fig.~\ref{mapcgran} the resulting Compton-Getting anisotropies when including the random striated component (left panel) and for the complete field including also the random isotropic component (right panel), for the JF12 model and for a rigidity of $E/Z =10^{17.8}$\,eV.  Note that the amplitude of the dipolar component drops to $d\simeq 0.0016$ when including the striated field and $d\simeq 7 \times 10^{-4}$ for the complete field case.\footnote{For simplicity we do not display  the maps for the JF+Planck model, but the features are qualitatively similar to those shown for  the JF12 model.} 

\begin{figure}[h]
    \centering
    \includegraphics[width=0.49\textwidth]{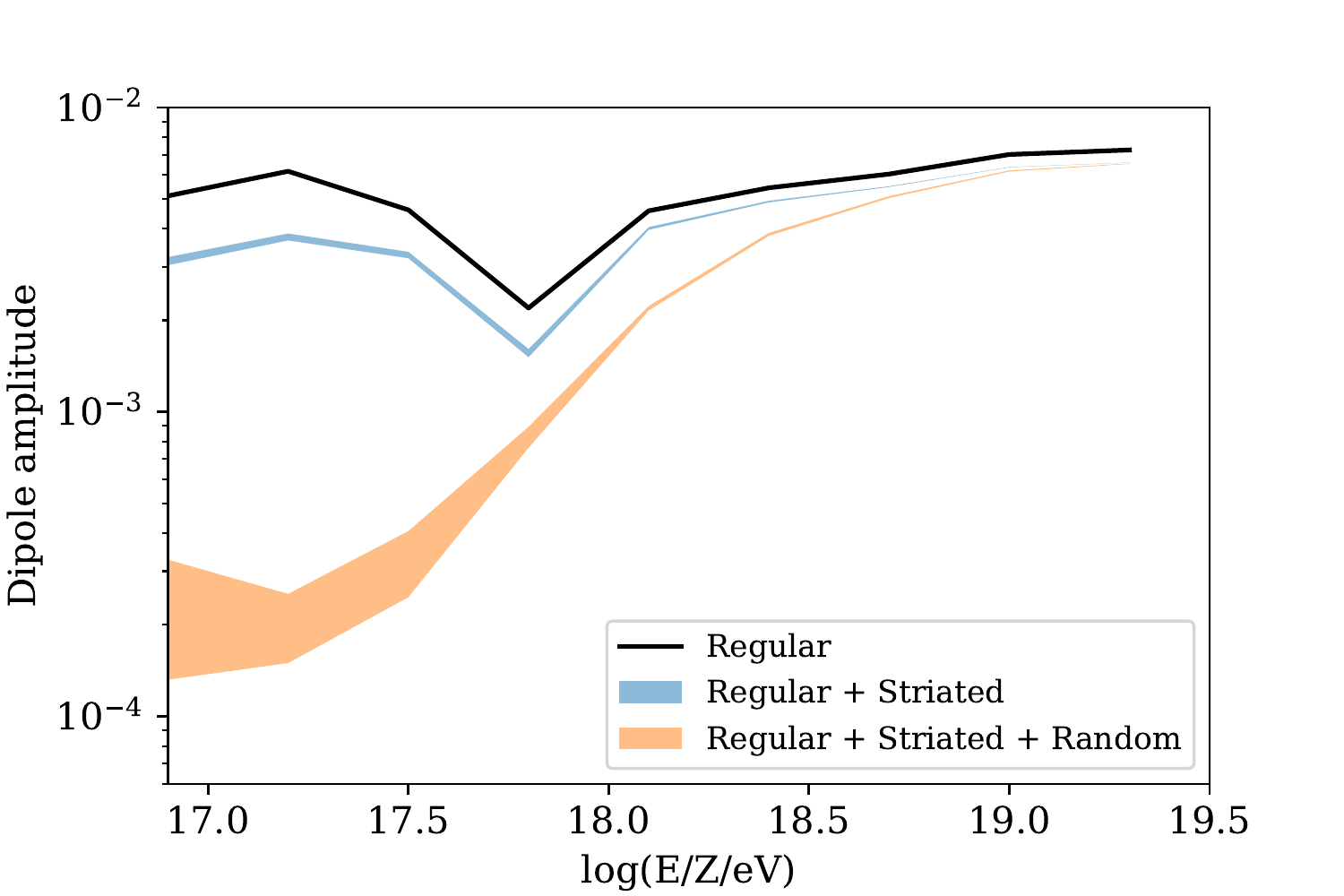} 
    \includegraphics[width=0.49\textwidth]{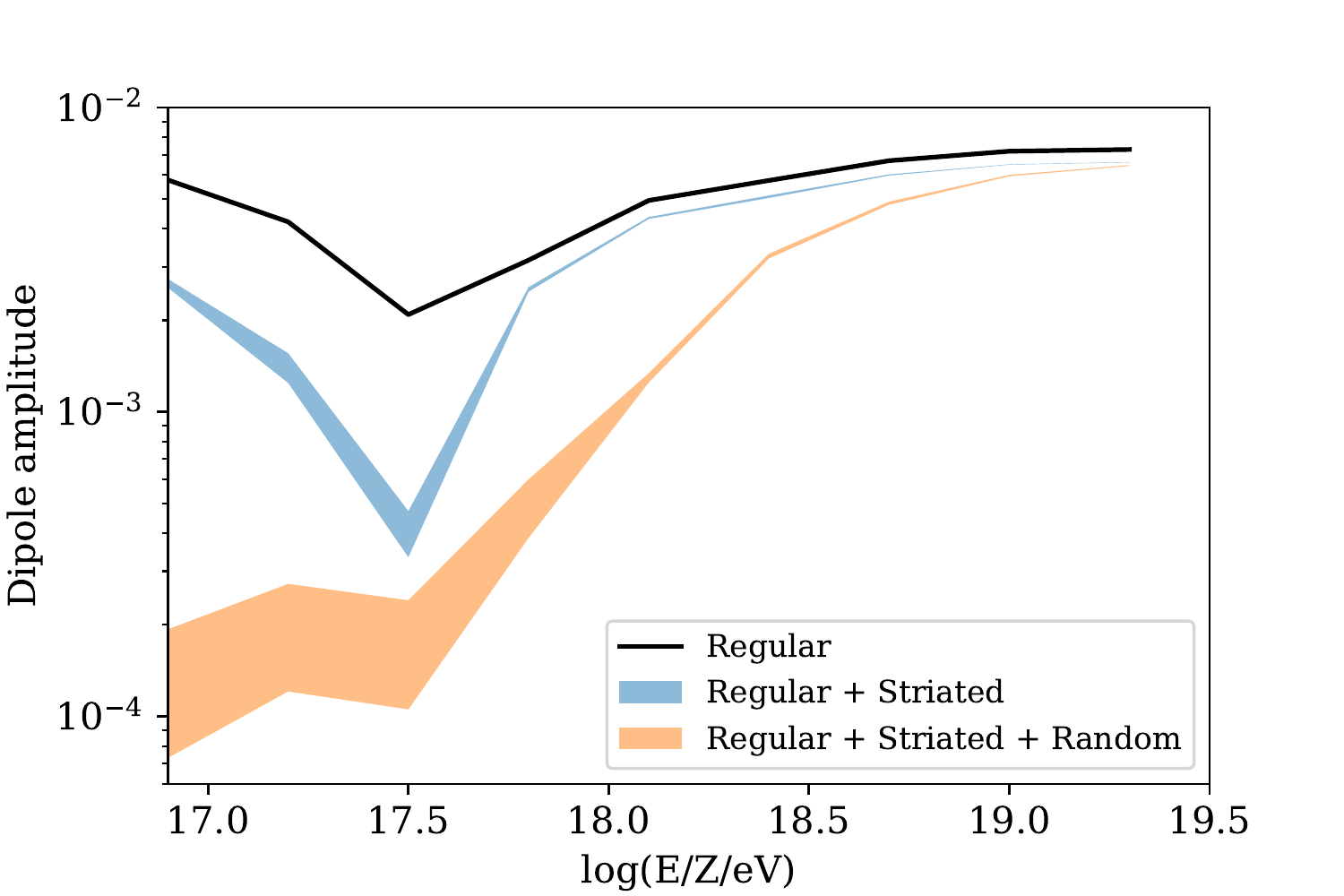} 
    \caption{Dipole amplitude resulting from the Compton-Getting effect after traversing the Galactic magnetic field as a function of  log($E/Z)$. The left panel corresponds to the magnetic field JF12 while the right panel corresponds to JF+Planck.}
    \label{ampvsE_CG}
\end{figure}

\begin{figure}[h]
    \centering
    \includegraphics[width=0.49\textwidth]{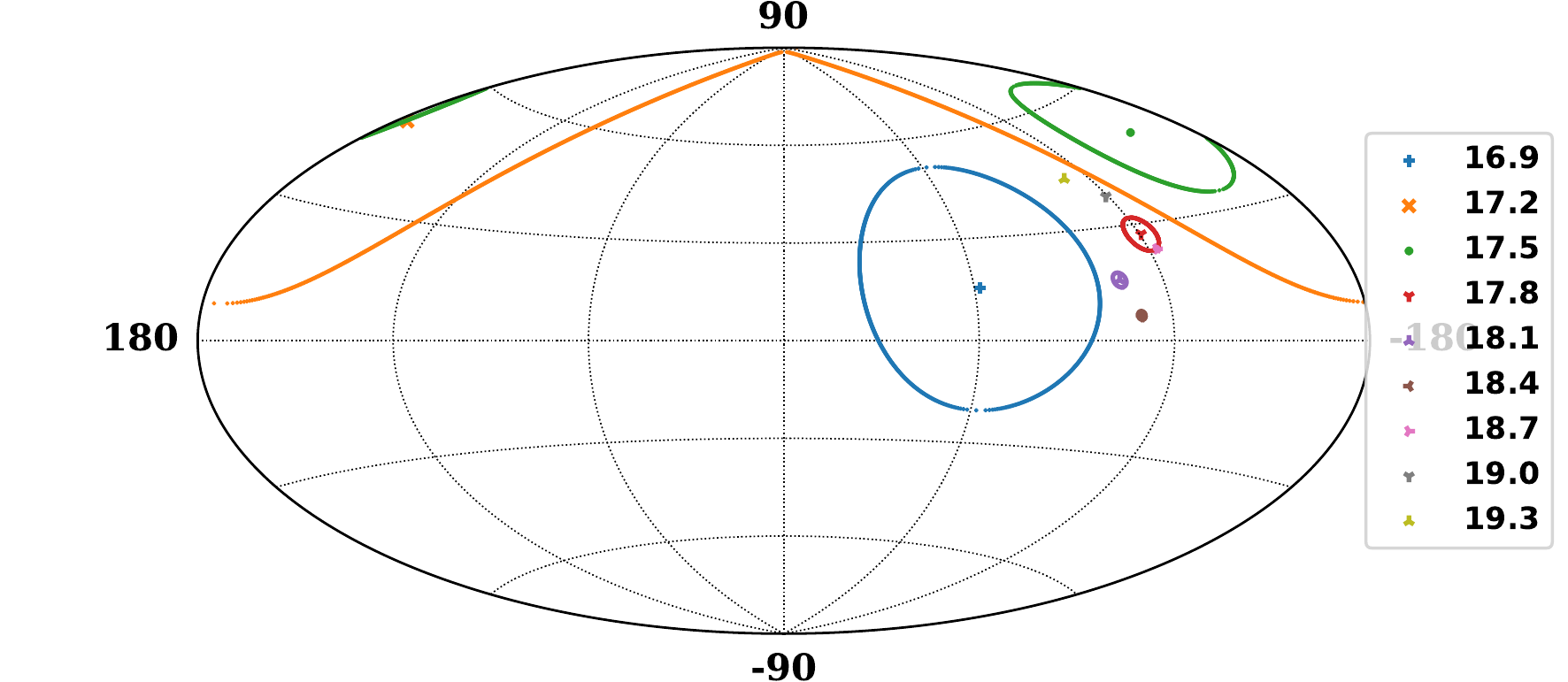} 
    \includegraphics[width=0.49\textwidth]{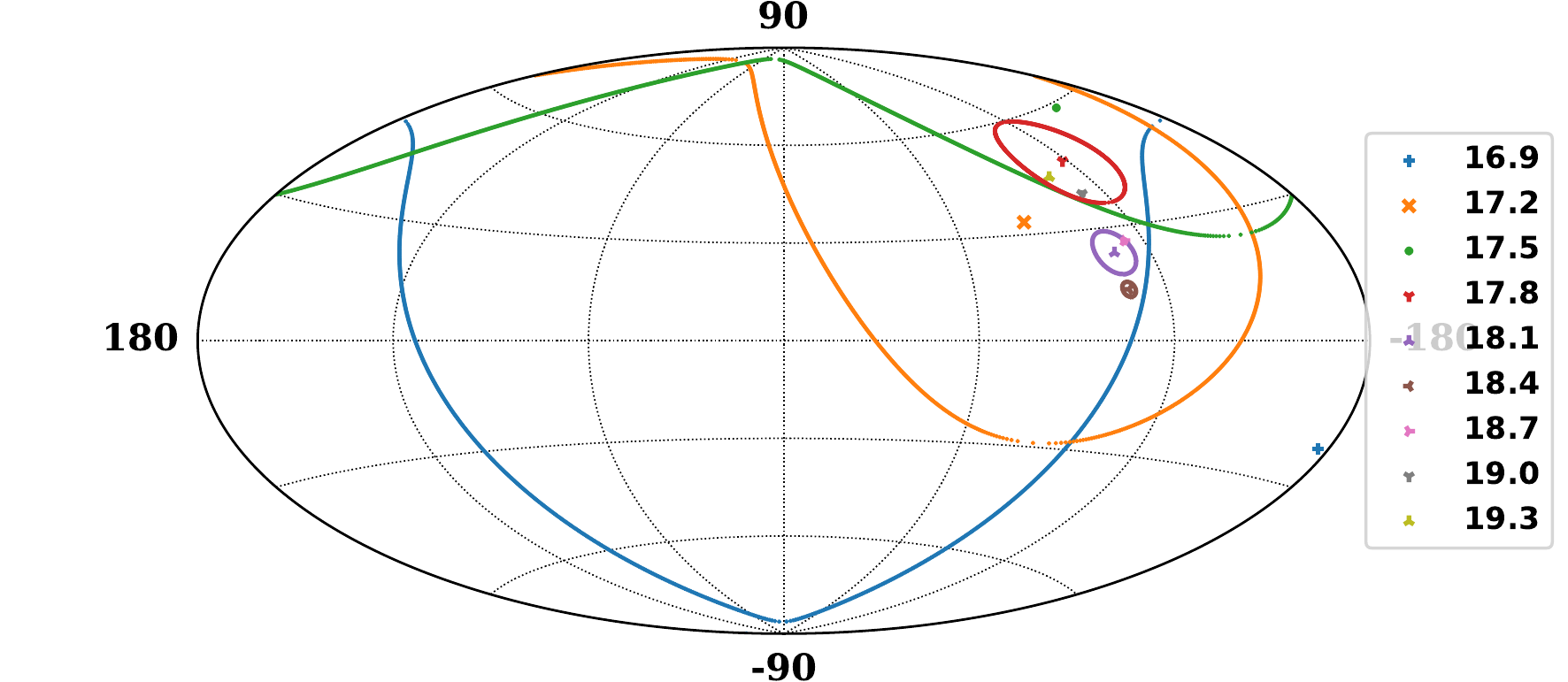}
    \caption{Maps in Galactic coordinates showing the average dipole direction from the Compton Getting effect for different values of log($E/Z/{\rm eV})$ when the complete magnetic field is considered. The circular contours indicate the median of  the simulated dipole directions. Left panel corresponds to the magnetic field JF12, right panel corresponds to JF+Planck. }
    \label{dirvsE_CG}
\end{figure}

In Fig.~\ref{ampvsE_CG} we show the amplitude of the dipolar component as a function of the rigidity $E/Z$, obtained when considering only the regular field, the regular and random striated components and the complete field including also the random isotropic component.  For the cases including random fields, the colored bands correspond to the dispersion around the mean values of ten simulations performed for each energy. At the highest energies the amplitude is close to that expected in the absence of magnetic field and for $\gamma=3.3$, which is $d \simeq 0.0066$, and then decreases for decreasing energies. When only the regular field or regular plus striated fields are included, the amplitude rises  at the lowest rigidities considered.
The inclusion of the random isotropic component leads to a smaller dipolar component, especially at the lowest rigidities considered, for both the JF12 (left panel) and JF+Planck (right panel) models. 

Figure~\ref{dirvsE_CG} shows the average dipole direction as well as the circular region around it within which the dipolar component points in half of the simulations, for different rigidities and for the case where the complete field  (regular, striated and random isotropic) is considered. At the highest energy the dipole  points close to the CMB dipole direction, while when the energy decreases  it moves down towards the Galactic plane, and then at the lowest energies it moves up again, with an increasing dispersion being obtained between the different realizations of the random field for decreasing energies.

\begin{figure}[h]
    \centering
    \includegraphics[width=0.49\textwidth]{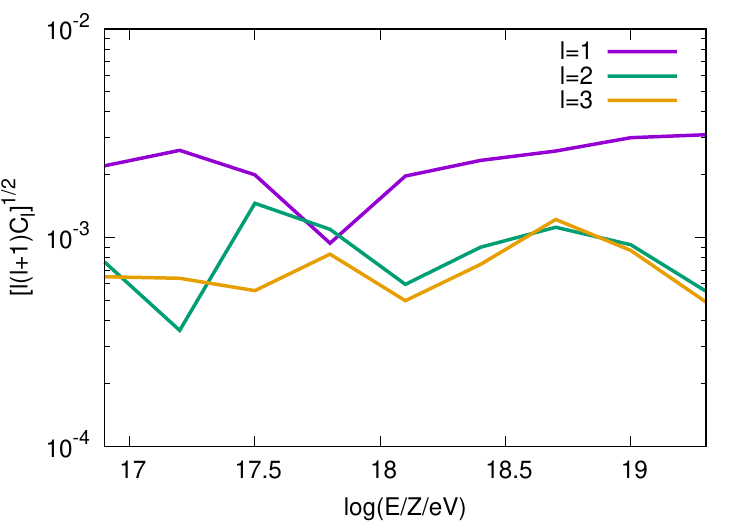} 
   \includegraphics[width=0.49\textwidth]{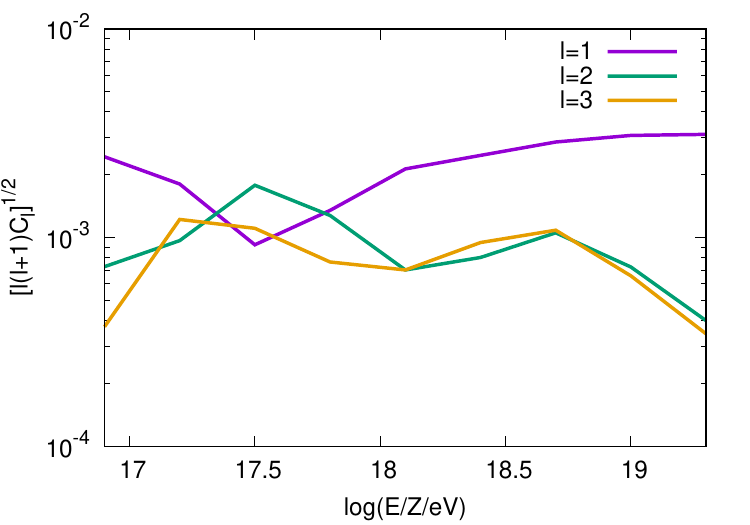} 
   
    \caption{Normalized multipole amplitude as a function of the rigidity for the three lowest $\ell$ values, assuming that the CRs are isotropic in the CMB rest frame. Left panel is  for the regular component of the JF12  model and right panel for that of the JF+Planck  model.}
    \label{cgmultipoles}
    \end{figure}

The deflections produced by the Galactic magnetic field not only change the direction and magnitude of the dipole present outside the halo, but they also produce significant distortions in the angular distribution pattern of the flux, as  is clearly seen in Figs.~\ref{mapcgreg} and \ref{mapcgran}. In order to quantify this effect, we show in Fig.~\ref{cgmultipoles} the normalized multipoles of order $\ell =1,\, 2$ and 3 as a function of the rigidity and considering just the regular component of the JF12 or JF+Planck models. The amplitude of the dipolar component results dominant with respect to the quadrupolar and octupolar ones both at the highest and lowest rigidities considered, while for  rigidities around $E/Z = 10^{17.5}$~eV the quadrupolar and  octupolar amplitudes become comparable to, or even larger than,  the dipolar one. This rigidity range is where the flip in the direction of the dipolar component occurs.

\begin{figure}[h]
\centering
\includegraphics[width=0.49\textwidth]{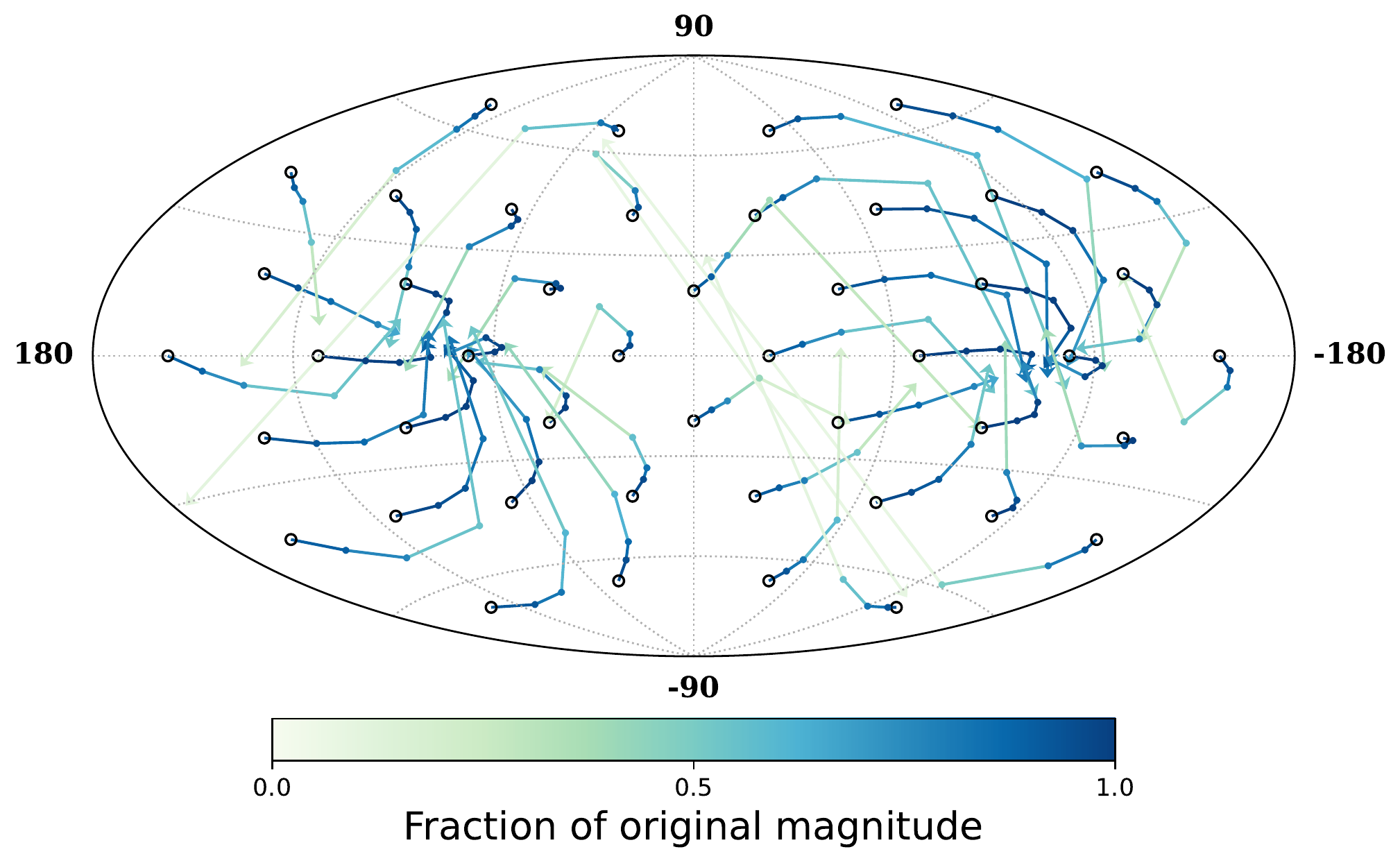} 
\includegraphics[width=0.49\textwidth]{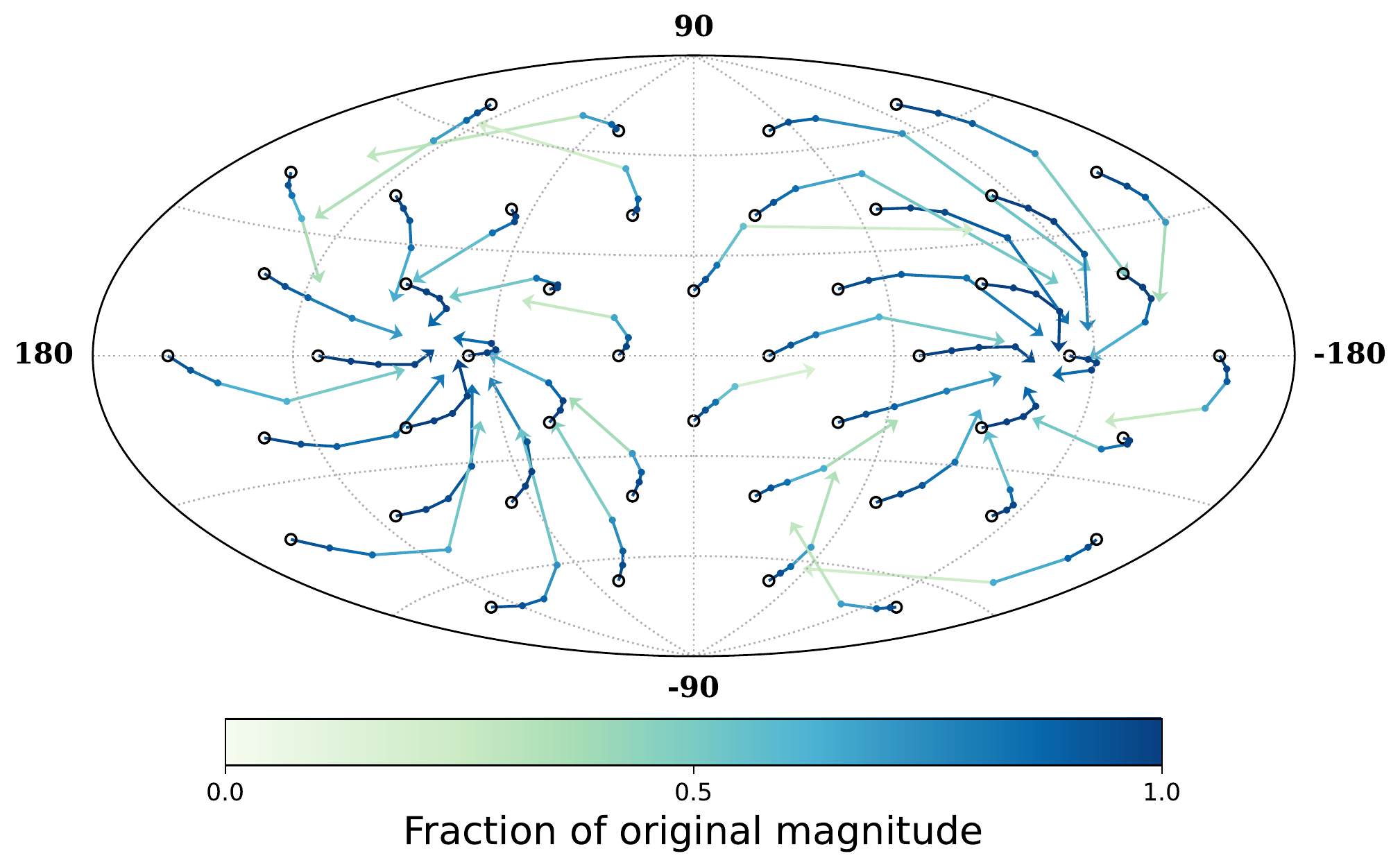} 
\includegraphics[width=0.49\textwidth]{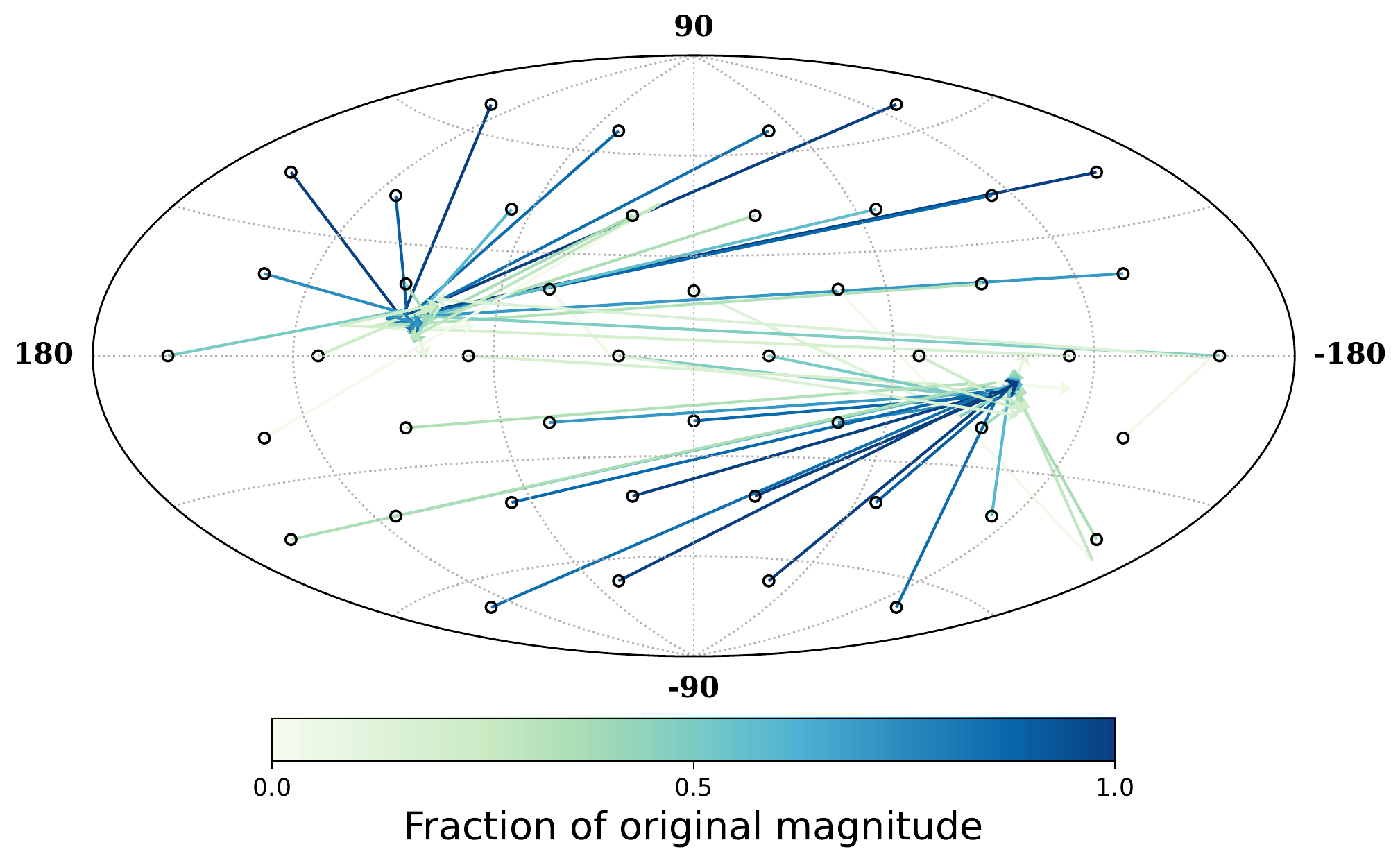} 
\includegraphics[width=0.49\textwidth]{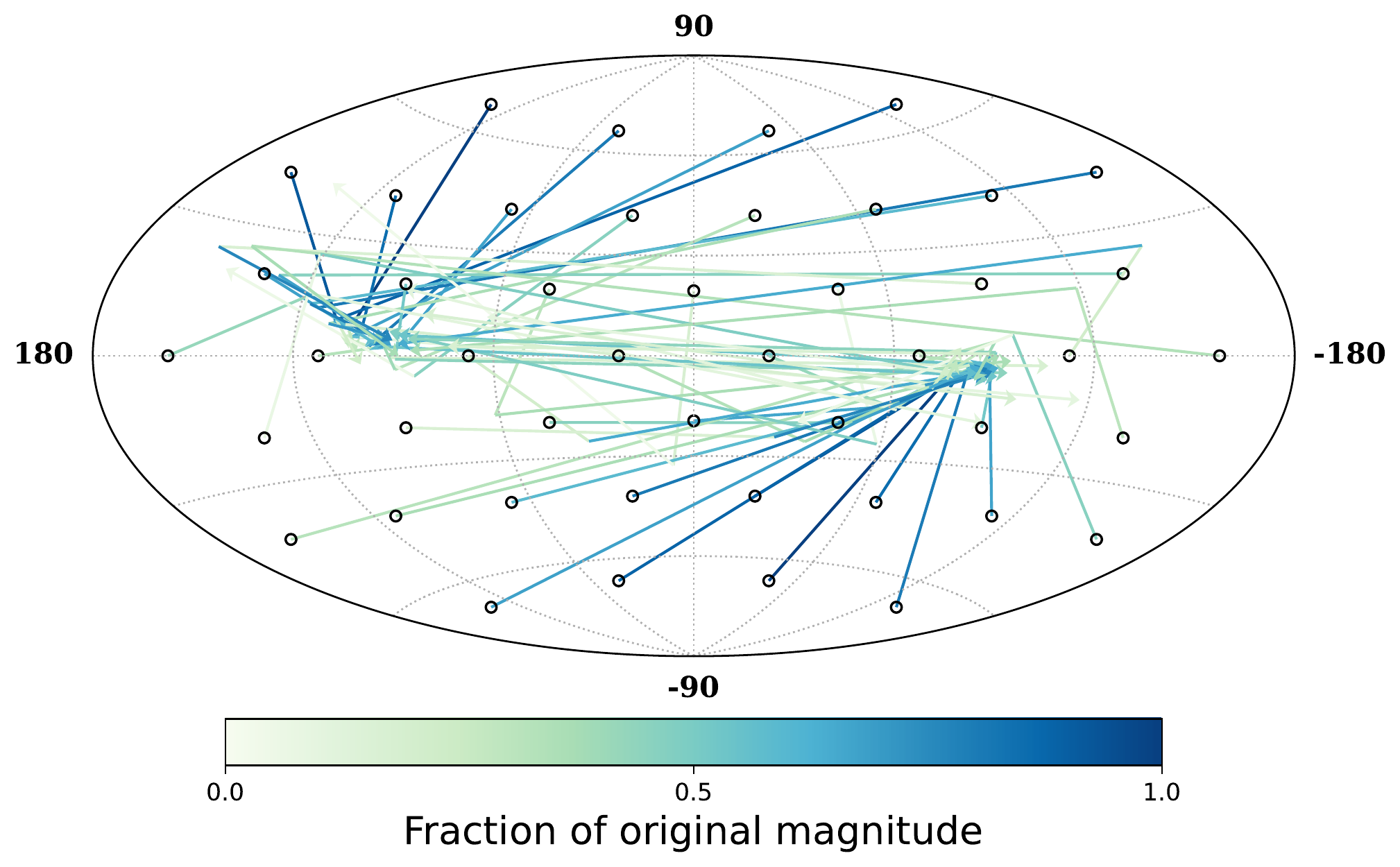} 

    \caption{Direction in Galactic coordinates of the dipolar component observed at the Earth for different dipolar directions outside the Galactic halo, after propagation in the regular component of the JF12 (left panel) and the JF+Planck (right panel)   magnetic field models. Black circles indicate the original dipole directions considered while the intermediate points along the lines indicate,  in the upper panels, the  direction at Earth for rigidities $E/Z = 10^{19.3}$~eV, $10^{19}$~eV, $10^{18.7}$~eV and $ 10^{18.4}$~eV, this last corresponding to the tip of the arrow. In the lower panels the intermediate point corresponds to $E/Z = 10^{17.2}$~eV, while the tip corresponds to $E/Z = 10^{16.9}$~eV. The color indicates the relative change in the amplitude of the dipole at Earth with respect to that outside the halo. }
    \label{dipdirs}
    \end{figure}
      
If the reference frame in which the CRs are isotropic were different from the CMB one, the solar system relative velocity $\vec V_{\rm iso}$ in Eq.~(\ref{phiearth}) would be different. The initial dipolar anisotropy, before being distorted by the magnetic field deflections, would in this case thus point towards a different direction and would have a different amplitude.  The modification of the direction and amplitude of the dipole by the effects of the Galactic magnetic field depends on its initial direction. We summarize the effect produced by the magnetic deflections, for different CR rigidities and initial directions of the maximum of the flux outside the halo, in Fig.~\ref{dipdirs}. The open circles correspond to the set of the original dipole directions considered, and the successive points  along the arrows indicate the direction of the corresponding dipolar component of the flux reaching the Earth for different CR rigidities, as described in the caption.
For rigidities above  $ {\rm few}\times 10^{18}$~eV, the  dipoles with originally positive Galactic longitudes tend to be rotated towards directions closer to the inner spiral arm, at $(\ell,b)\simeq (80^\circ,0^\circ$), while those with original negative longitudes tend to be rotated towards the outer spiral arm direction, at $(\ell,b) \simeq (-100^\circ,0^\circ)$, as  is shown in the upper panels in Fig.~\ref{dipdirs}. On the other hand, for rigidities below  $ {\rm few}\times 10^{17}$~eV, the dipoles with originally positive Galactic latitudes tend to be rotated towards  directions closer to the inner spiral arm, while those with originally negative latitudes tend to be rotated towards the outer spiral arm direction, as  is shown in the lower panels of the figure. We then see that the change in the dipolar phase from negative Galactic longitudes at high energies to positive ones at low energies, obtained before  for the case in which the dipolar direction coincides with the CMB one, is actually expected to happen more generically for original dipole directions roughly in the quadrant with negative longitudes and positive latitudes. On the other hand, no drastic change of phase is expected to appear for original dipole directions lying in the quadrants with positive latitudes and longitudes or negative latitudes and longitudes. Finally, if the original dipole direction were in the  quadrant with positive longitudes and negative latitudes, the dipole at Earth is expected to display a change from  directions closer to the inner spiral arm at high energies to directions closer to the outer spiral arm at low energies. 

If the CR flux  reaching the Galactic halo were to have a significant intrinsic dipolar anisotropy, as for example resulting from the diffusion in the extragalactic magnetic field of CRs from some nearby powerful sources, this anisotropy would be affected by the deflections in the Galactic magnetic field in the same way as  was just discussed  for a Compton-Getting dipole. As a result, the direction and the amplitude of the dipole observed at the Earth would differ from the values outside the halo in a way similar to what is shown in Fig.~\ref{dipdirs}. 

\section{Combined results}

In the previous sections we have discussed the effects of the Galactic magnetic field upon the distribution of extragalactic CR arrival directions, considering separately the acceleration due to the electric force arising from the Galactic rotation and the modification of a Compton-Getting term due to magnetic deflections. The former effect is expected to give the dominant contribution at the lower rigidities. The latter one depends on the relative velocity of the observer with respect to the  reference frame in which CRs are approximately  isotropic,  which is however unknown, but for definiteness we will consider in this discussion the plausible case described in detail before in which this frame coincides with the CMB one.

\begin{figure}[ht]
    \centering
    \includegraphics[width=0.49\textwidth]{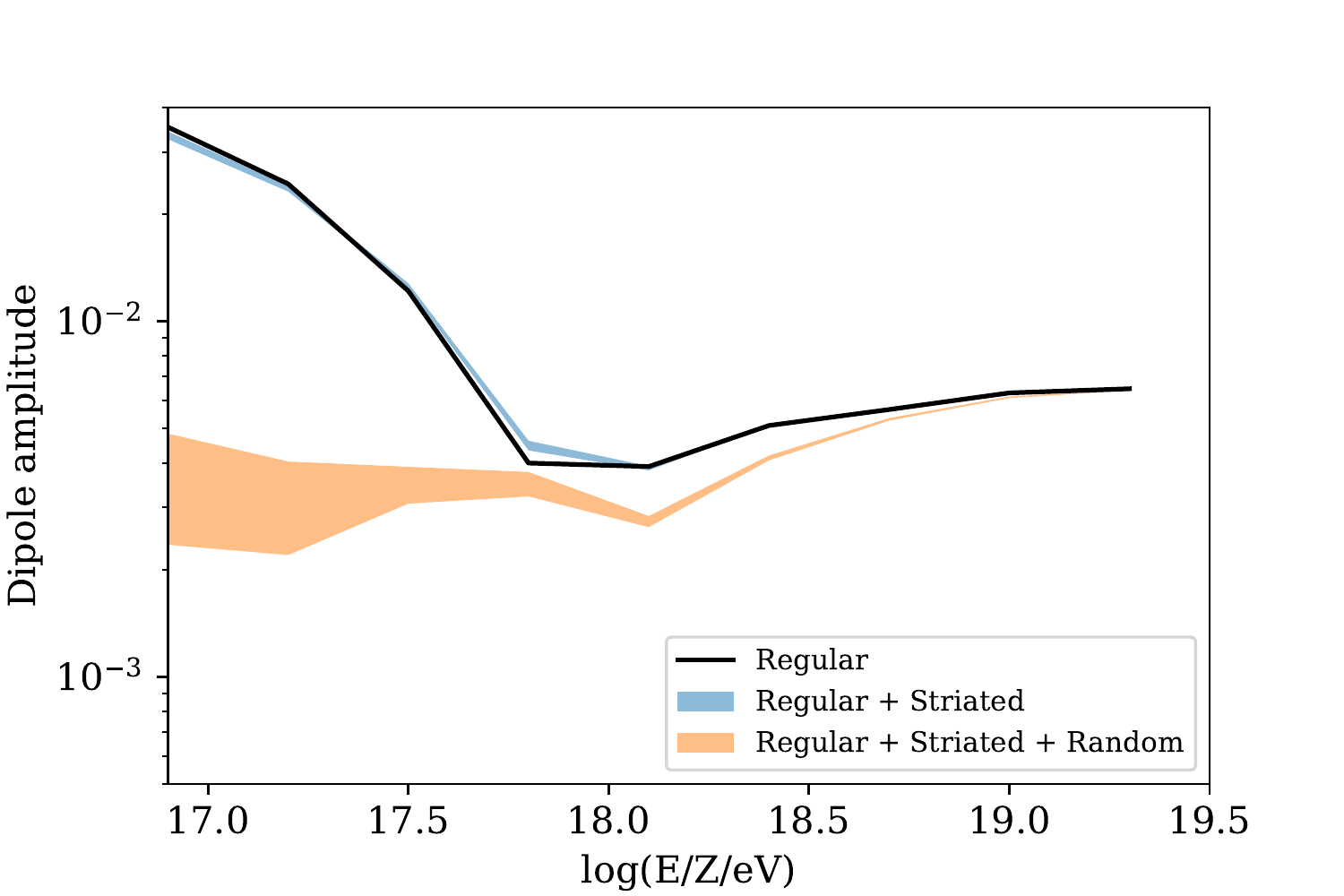} 
    \includegraphics[width=0.49\textwidth]{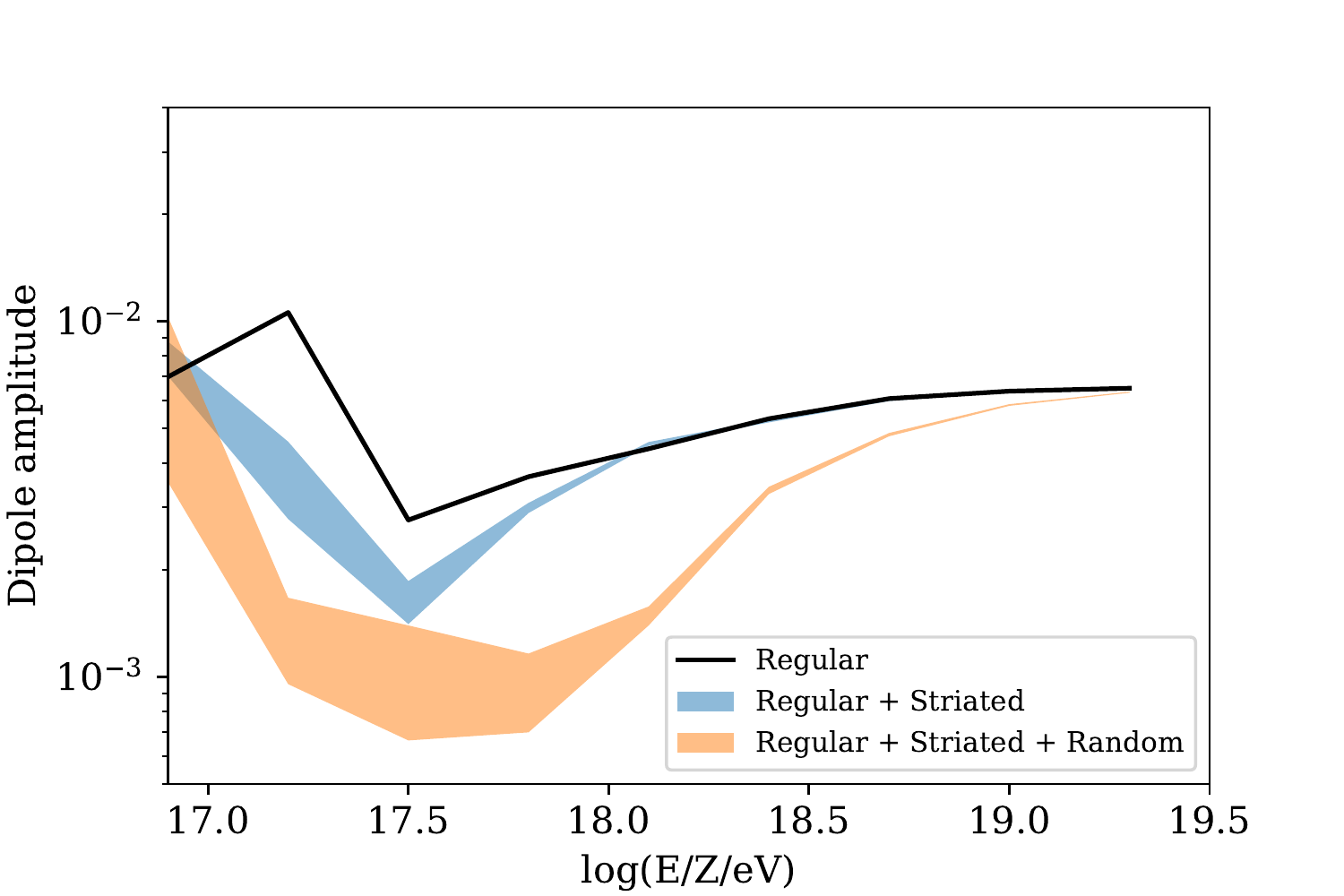} 
    \caption{Total dipole amplitude resulting from both the acceleration due to electric field associated with the Galactic rotation and the Compton-Getting effect, after traversing the Galactic magnetic field and as a function of the logarithm of the rigidity $E/Z$. The left panel corresponds to the magnetic field model JF12,  the right panel corresponds to JF+Planck model.}
    \label{ampvsE_Tot}
\end{figure}

We show in Fig.~\ref{ampvsE_Tot} the total dipolar amplitude resulting from the combination of both effects, as a function of $E/Z$, for the  cases in which just the regular magnetic field component is considered, the case in which the striated field is also included and for the complete magnetic field with the random isotropic component, both for the JF12 and JF+Planck models. At the highest energies the amplitude tends to the value expected from the Compton-Getting effect, $d \simeq 0.0066$, and it then decreases for decreasing energies as was  discussed in Section \ref{comptongetting}. However, when looking at the regular or regular plus random striated cases, we see that for rigidities below $E/Z \sim 10^{18}$~eV (for the JF12 model) or $E/Z \sim 10^{17.5}$~eV (for the JF+Planck model) the amplitude rises again for decreasing energies. This rise in the amplitude results from the acceleration due to the electric force. As  was discussed in Section~\ref{acceleration}, the acceleration effect is larger in the JF12 model than in the JF+Planck and thus the crossover of the two effects occurs at higher rigidities in the JF12 case.  When the random isotropic magnetic field component is also included, the amplitude for the lower rigidities gets significantly suppressed.

\begin{figure}[ht]
    \centering
    \includegraphics[width=0.49\textwidth]{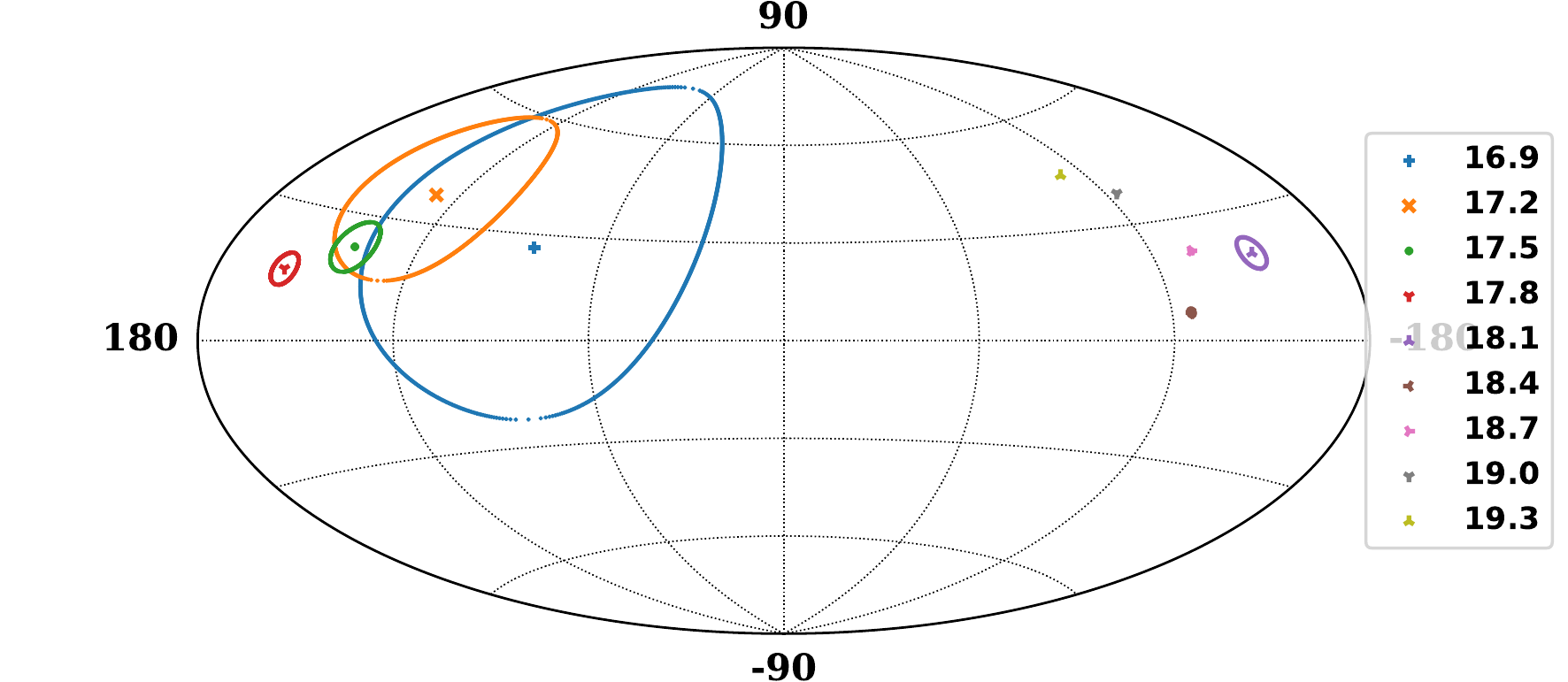} 
    \includegraphics[width=0.49\textwidth]{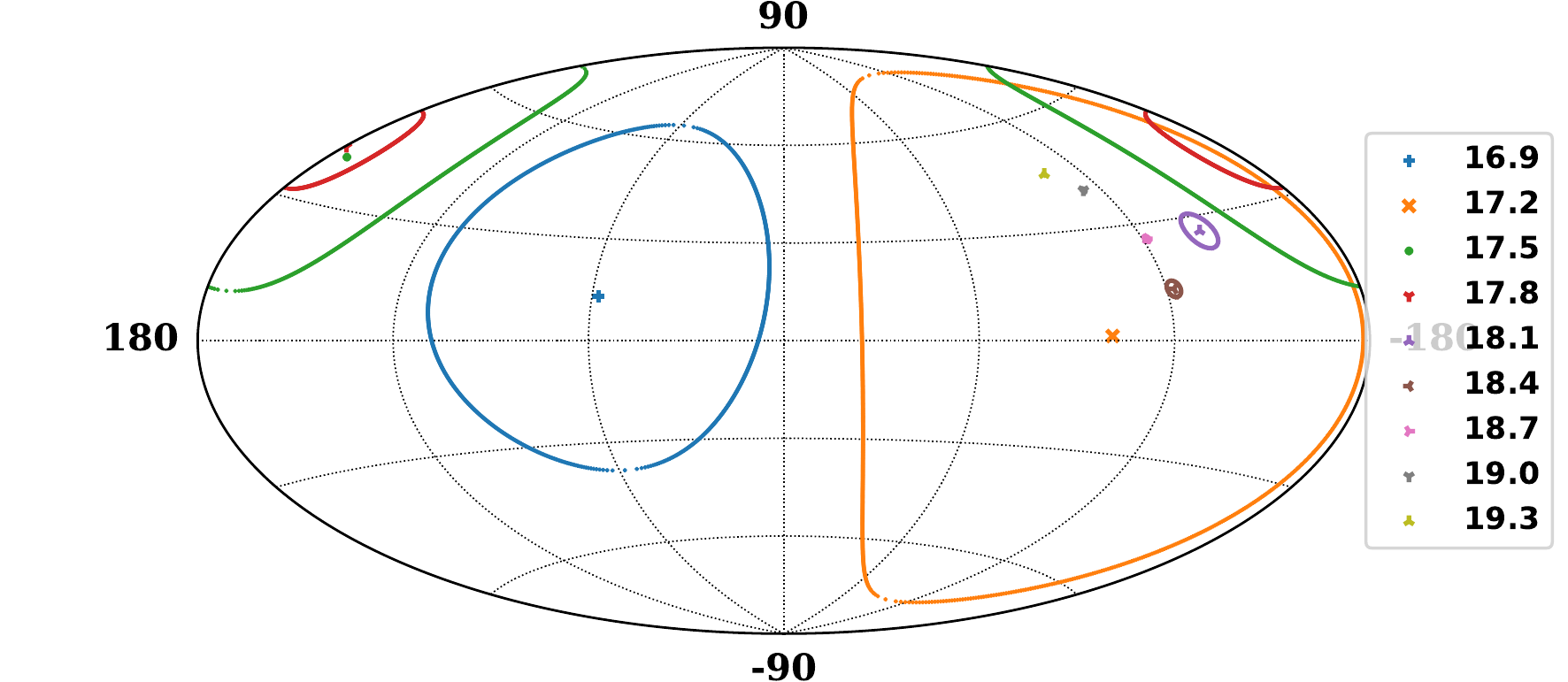}
    \caption{Maps in Galactic coordinates showing the average total dipole direction for different values of log($E/Z$/eV) when the complete magnetic field is considered.  The circular contours indicate the median of the dipole's direction distribution. Left panel corresponds to the magnetic field JF12, right panel corresponds to JF+Planck.}
    \label{dirvsE_Tot}
\end{figure}

Figure~\ref{dirvsE_Tot} shows the corresponding dipole directions in the sky for the case of the complete field and for both magnetic field models. A significant change in the dipole direction is apparent, specially  for the JF12 model, from positive Galactic longitudes for  rigidities  below $\sim 1 $~EeV to negative Galactic longitudes for larger rigidities, corresponding respectively to the predominance of the acceleration or the Compton-Getting effects.

Since the anisotropies resulting from the two effects discussed in this paper depend on the rigidity of the particles, $E/Z$, if the flux of particles reaching the Earth has a mixed composition the anisotropy at a given energy will result from the superposition of the anisotropies corresponding to the different rigidities associated with each mass component. The Pierre Auger Observatory  measurements indicate that the composition for energies above $10^{17.2}$~eV  has indeed a mixed nature \cite{be17}. In particular, for energies below the ankle a mixed composition of protons and intermediate mass nuclei is inferred.  We will hence consider, as a simplified approximation,  that in this energy range there is a constant fraction of 50\% of H and 50\% of N, both of extragalactic origin,  with a negligible contribution from the Galactic component. 

\begin{figure}[h]
    \centering
    \includegraphics[width=0.49\textwidth]{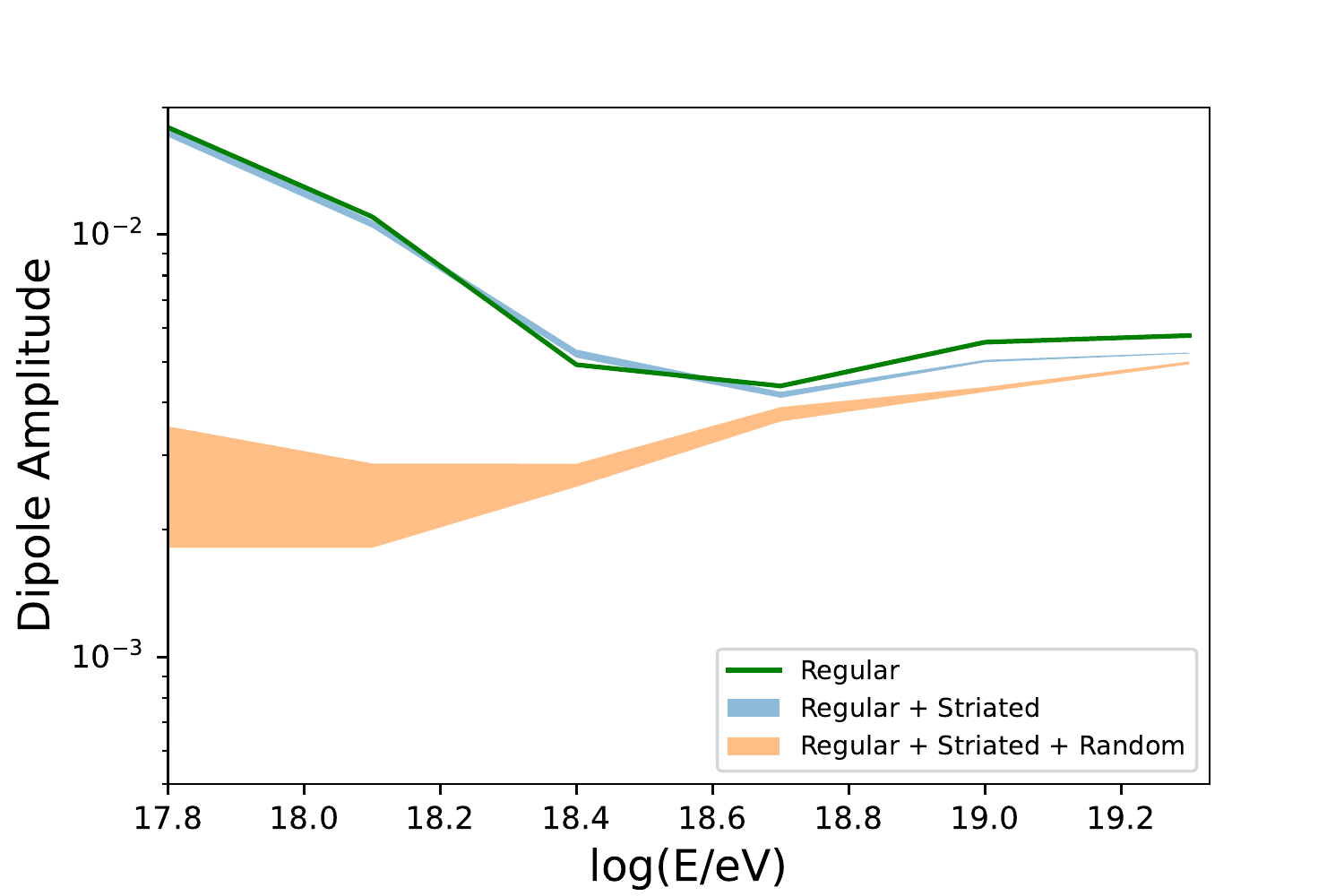} 
    \includegraphics[width=0.49\textwidth]{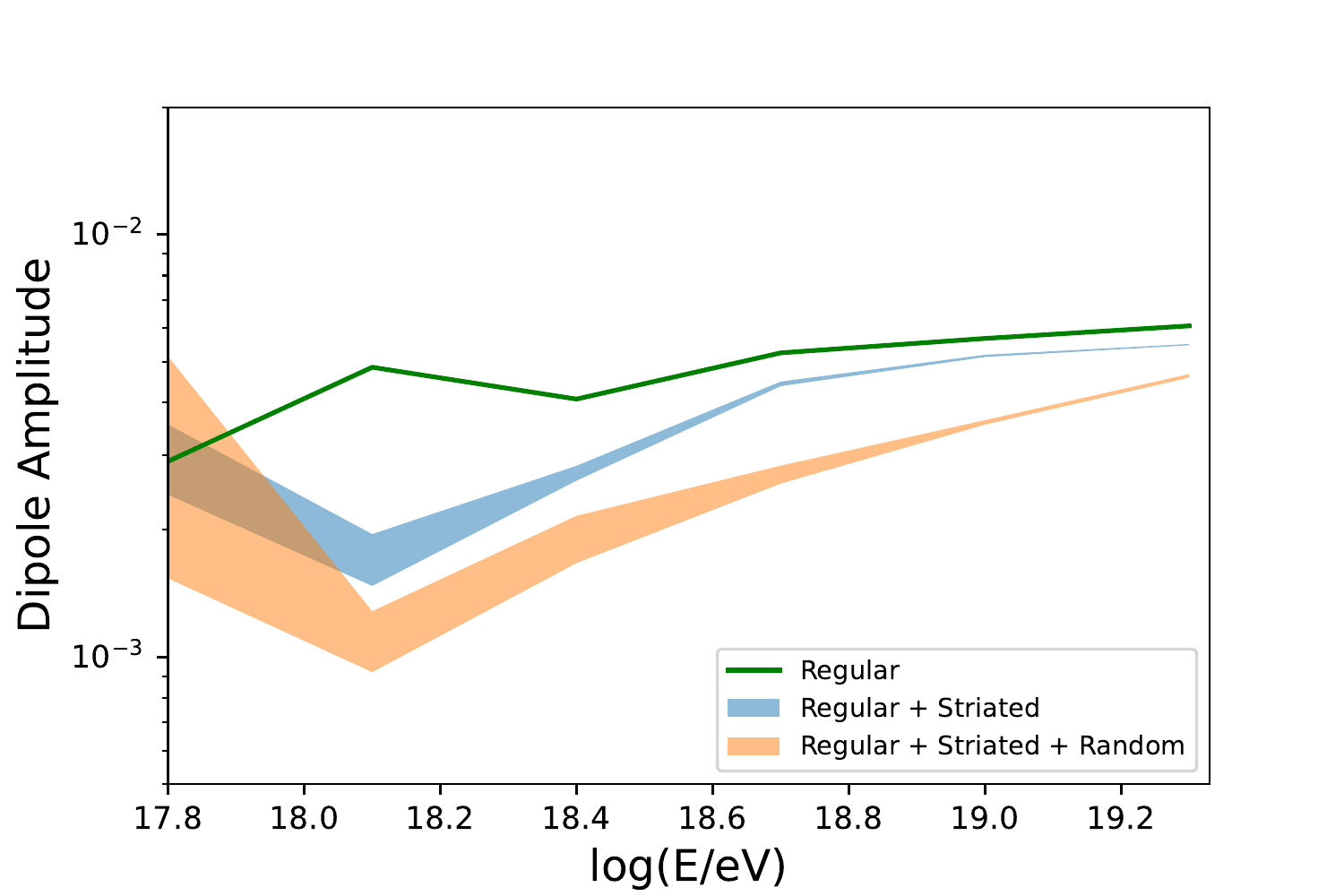}
    \caption{Total dipole amplitude as a function of the energy in a mixed composition model (50\% of H and 50\% of N)
    of extragalactic cosmic rays assuming that they are isotropic in the CMB rest frame. Left panel corresponds to the magnetic field JF12, right panel corresponds to JF+Planck.}
    \label{dvsE_mixed}
\end{figure}

For a given mixture of elements in the flux of cosmic rays reaching the Galaxy, we can compute the expected dipolar amplitude of the anisotropy at the Earth for each energy $E$ by adding the dipole vectors for the corresponding rigidities $E/Z$ weighted by the fraction of particles with each charge $Z$. We show in Fig.~\ref{dvsE_mixed} the total dipolar amplitude as a function of the energy in the mixed composition model for the JF12 (left panel) and for the JF+Planck (right panel) models. For high energies the amplitude tends to the Compton-Getting value ($d \simeq 0.0066$) in both models, since CRs are not much affected by the Galactic magnetic field. A more noticeable difference between the two models appears at low energies, especially when looking at the regular field component, with a larger amplitude being obtained for the JF12 model. This is due to a partial cancellation at these energies between the anisotropies of the H and N components in the JF+Planck model.

 One should also note that at lower energies a transition to a regime dominated by the Galactic CRs should start to take place,  and hence approaching the second-knee energy the anisotropies should progressively tend to those of the Galactic component, not considered here. The fact that the effects of the random magnetic fields become more notorious at low rigidities is also indicating that a transition to a diffusive regime inside the Galaxy is taking place for the incident extragalactic CRs. 

\section{Discussion and conclusions}

The large angular scale  modulation of the flux of cosmic rays  can be better measured as a function of right ascension (rather than declination), since  taking advantage of the uniform rotation of the Earth the experimental spurious effects are better  controlled. From the observed modulation of the flux as a function of  right ascension it is possible to estimate the projection of the dipole in the plane orthogonal to the Earth rotation axis, $d_\perp$, and the phase of the maximum of this modulation corresponds to the right ascension of the dipole direction. We will then shift here to consider equatorial coordinates so as to allow for a better comparison with observations.

\begin{figure}[h]
    \centering
    \includegraphics[width=0.49\textwidth]{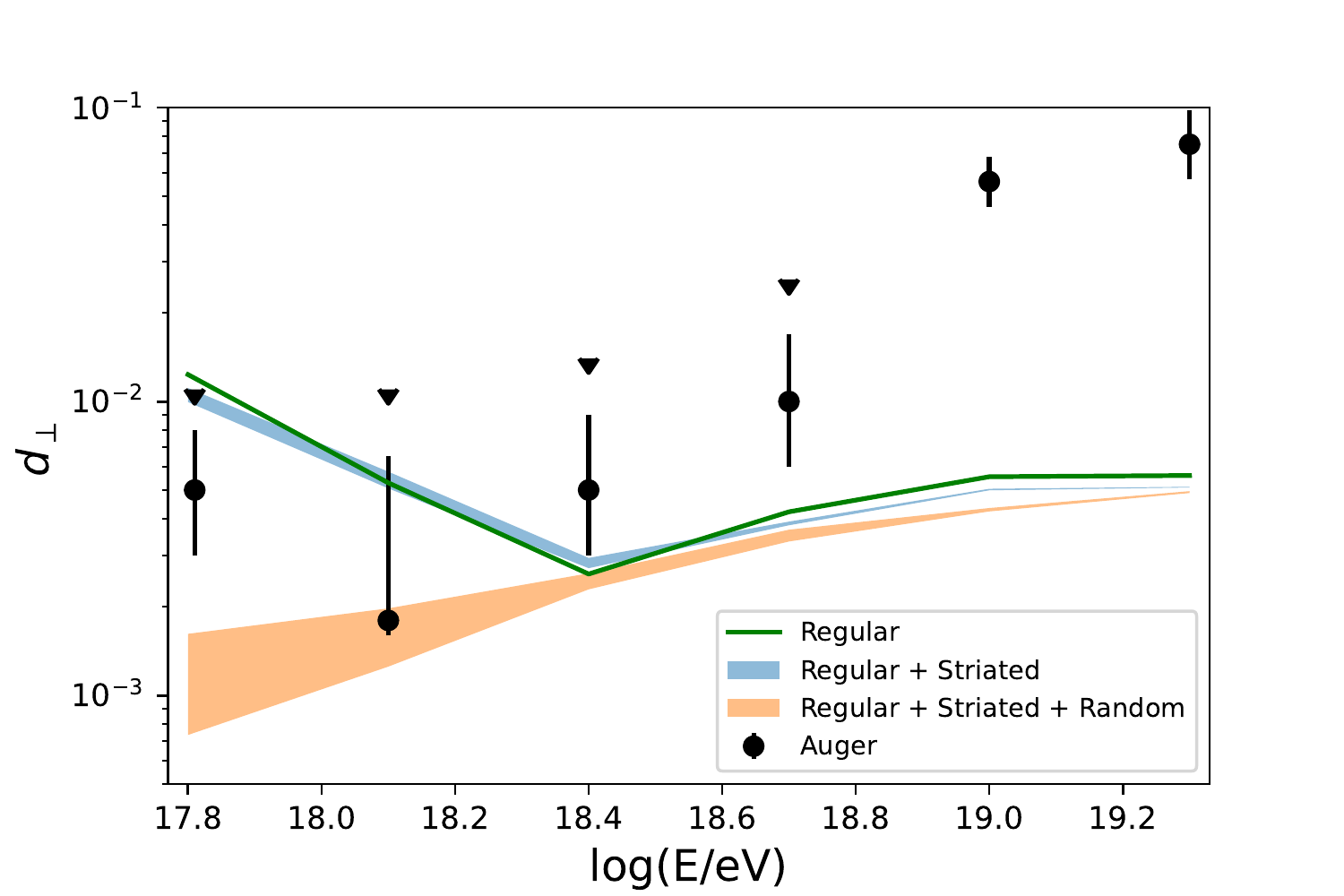} 
    \includegraphics[width=0.49\textwidth]{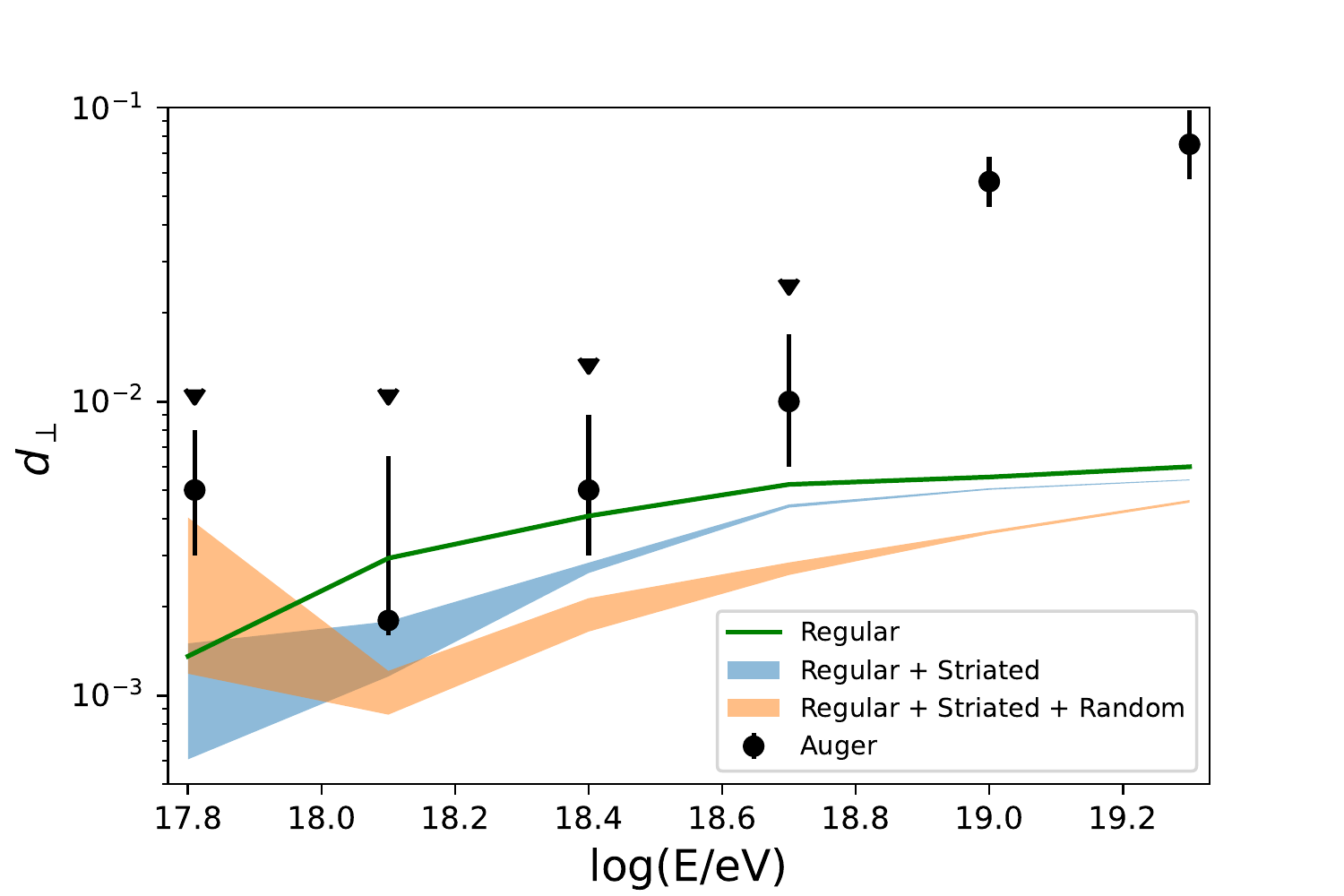}
        \includegraphics[width=0.49\textwidth]{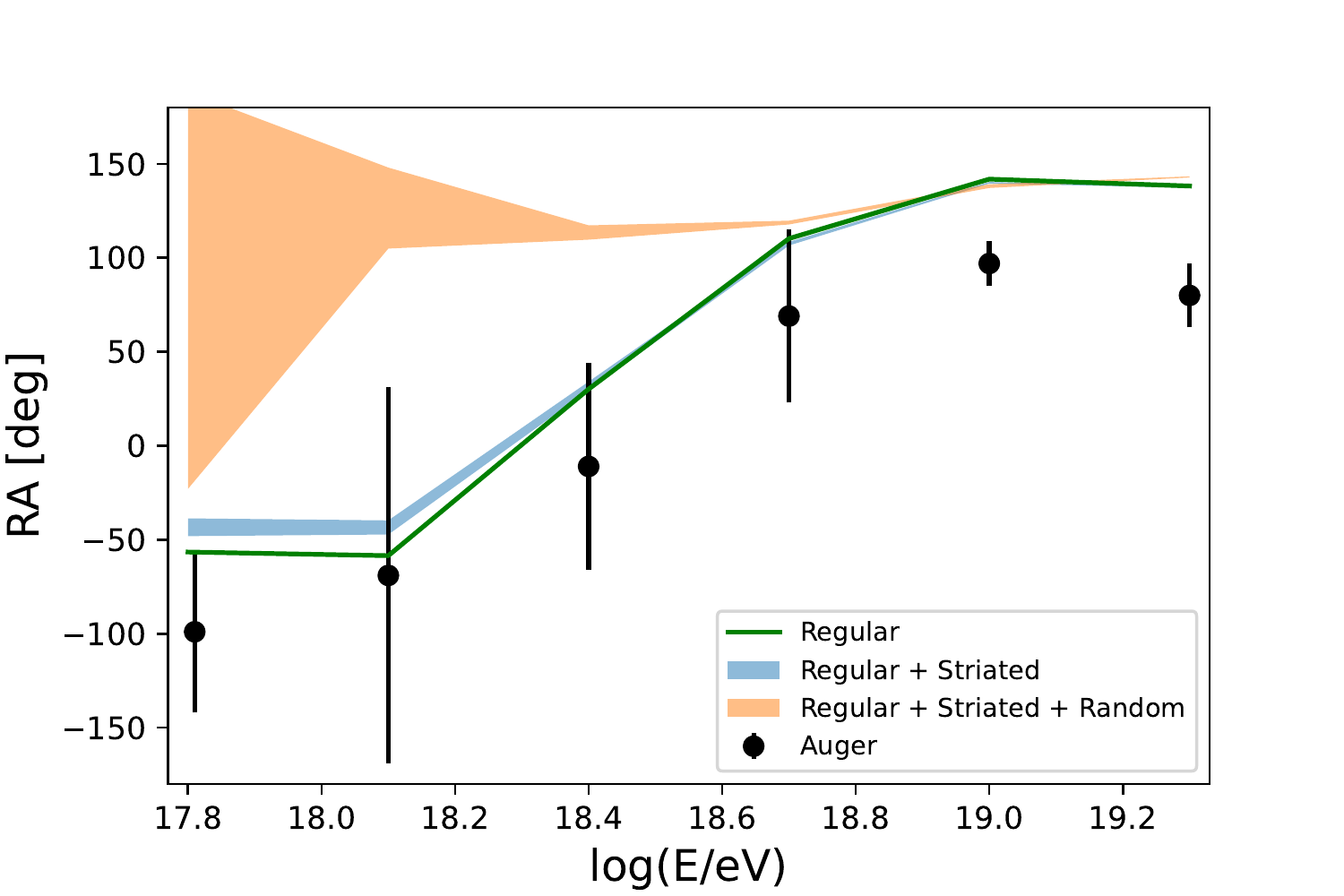} 
    \includegraphics[width=0.49\textwidth]{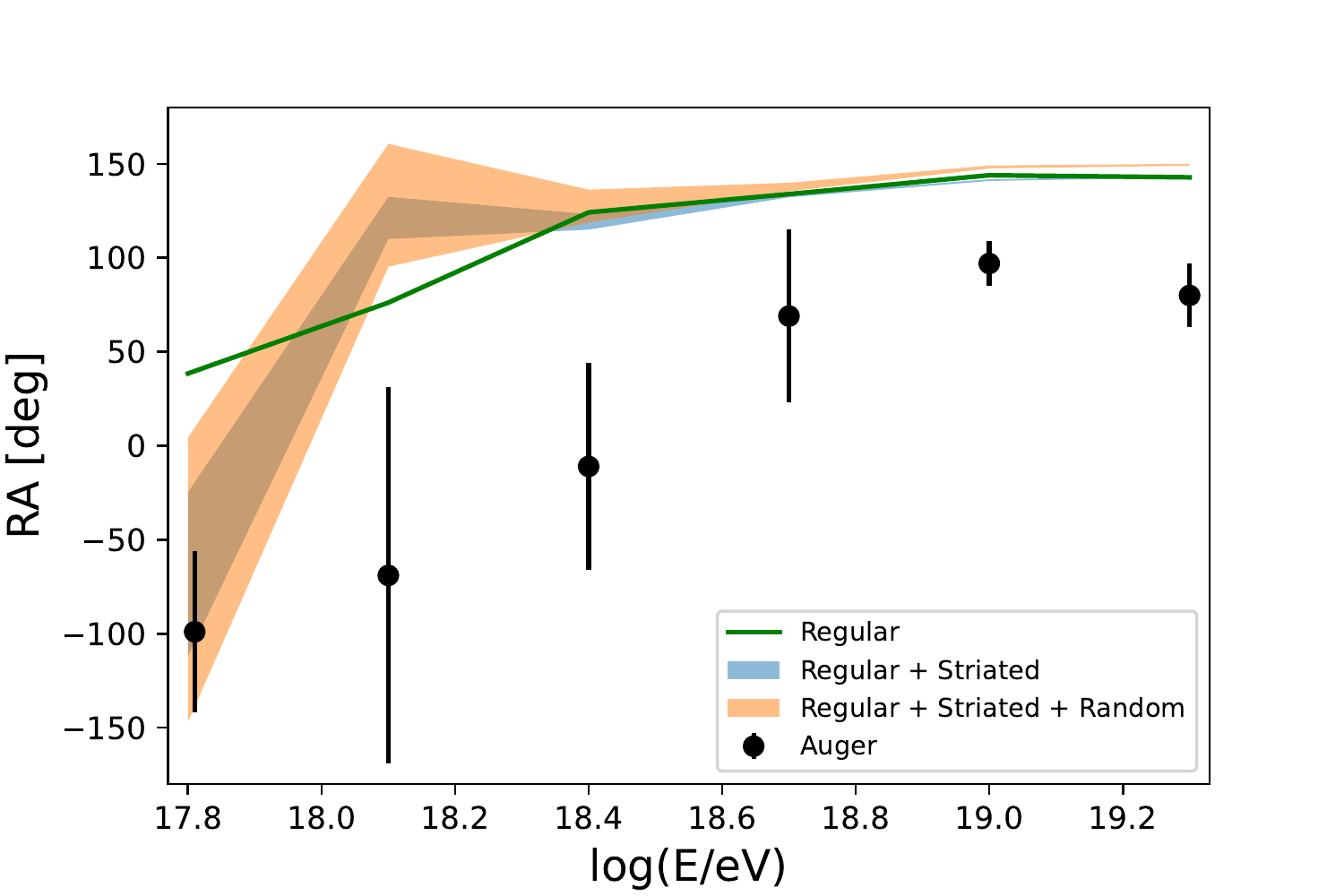}
    \caption{ Amplitude of the equatorial dipole (upper panels) and its right-ascension phase  (lower panels) in a mixed composition model (50$\%$ of H and 50$\%$ of N). Results are shown for the regular, regular + striated and complete magnetic field scenarios. The points with error bars are the measurements from the Pierre Auger Observatory \cite{lsa20}. For energies below 8~EeV, where the measurements have  chance probability larger than 1\%, the downward triangles in the $d_\perp$ plots indicate the 99\% CL upper limits obtained.  Left panels correspond to the magnetic field JF12, right panels correspond to JF+Planck.}
    \label{dperpvsE_mixed}
\end{figure}

We show in Fig.~\ref{dperpvsE_mixed} the amplitude of $d_\perp $ and its right-ascension phase as a function of the energy, as  expected for the JF12 (left panel) and the JF+Planck (right panel) models and considering only the regular component, the regular plus the striated components or the complete field with also the random isotropic component.
We also included in the plot the results of the measurements from the Pierre Auger Observatory \cite{lsa20}. 

The observed phase shows a transition from  $ {\rm RA} \simeq -90^\circ$ for energies below 1~EeV to  ${\rm RA}\simeq 90^\circ$ for energies above  4~EeV. Note that a very similar shift in the  phase as a function of energy is obtained from the effects of  the regular or the regular plus striated components of the JF12 model. However, the addition of the isotropic random component somewhat spoils this good agreement between model predictions and observations. It should be mentioned however  that several authors have pointed out that the comparison of the predictions of the original JF12 model for the synchrotron emission maps with the final WMAP data release, and also with data from Planck, indicates that a significant reduction of the assumed amplitude of the isotropic random component is required \cite{beck16,UF19}. A reduction in the isotropic random field amplitude is on the other hand  expected to lead to predictions for the dipolar components in between those obtained for the regular and for  the complete field models that were shown in Fig.~\ref{dperpvsE_mixed}, and hence should improve the agreement with the data.  One can also see that the transition in the phase occurs for lower energies in the JF+Planck model than in the JF12 one.

In the energy bins below 8~EeV  no significant measurement has been obtained for the equatorial dipole amplitude $d_\perp$, and  the 99\% CL upper bounds on the amplitude are also displayed in the plot for those bins. For both  magnetic field models considered the predicted amplitude is compatible with the observations. For energies around 10~EeV a significant dipolar component has been measured by the Pierre Auger Collaboration, with an amplitude of about $d \simeq 7\%$ \cite{LSAscience,LSICRC21}, increasing approximately linearly with energy above 4~EeV and up to the highest  energy bin considered \cite{LSA18,LSICRC21}. This anisotropy is expected to be associated with the sources of the highest energy CRs that dominate the flux for energies above the ankle. Since at the highest energies CRs are limited to arrive to the Earth from relatively closeby sources, lying at distances smaller than about 100~Mpc, a plausible explanation is that the observed anisotropy arises from the inhomogeneous distribution of the CR sources in the local Universe. The distribution of matter around us, as traced by galaxies in the 2MRS catalog, does in fact show a significant dipolar component with a maximum not far from the CMB dipole maximum  \cite{erdogdu}. This is in fact expected as it is the gravitational attraction from the nearby matter what produces the Local Group peculiar velocity that is ultimately responsible for the CMB dipole. The anisotropy of this high-energy component is thus expected to have a  direction not very different from the Compton-Getting one but with a larger amplitude, making it the dominant contribution above the ankle. This could in principle allow to also account for the measurements at the two highest energy points shown in Fig.~\ref{dperpvsE_mixed}.

 A significant uncertainty in this study is due to the poor knowledge of the Galactic magnetic field. Different observations, like Faraday rotation measures (RM) and the intensity of total and polarized diffuse synchrotron emission, are used to constrain different magnetic field components.\footnote{As described in \cite{jaffe10},  the regular field component contributes to both polarized emission as well as to RM, while the isotropic random component contributes only to the total intensity of the diffuse emission. The striated, or `random ordered', component contributes to the polarized emission but not to RM. } These observables depend on different projections of the magnetic field, along or perpendicular to the line of sight, and other not well known quantities like the electron density along the line of sight, what leads to different uncertainties affecting  the coherent, random striated and random isotropic components (for a state of the art review of Galactic magnetic field modelling see \cite{jaffe19}). For this reason, we have presented separately the results due to the regular component alone, for the regular and the striated components and for the complete field including also the random isotropic component.

As a summary,  we have explored in this paper the effects of the Galactic magnetic field on the large angular scale distribution of the flux of extragalactic cosmic rays reaching  Earth, focusing on the energy range between the second knee and the ankle. In this energy range anisotropies leading to dipolar amplitudes at the few per mille level are predicted both from the Compton-Getting effect and from the acceleration due to the electric force induced by the rotation of the Galaxy. These amplitudes are consistent with the existing bounds on the equatorial component of the dipole amplitude below the ankle energy, and may lead to a change in the right-ascension phase in a way compatible with the observations, depending on the details of the magnetic field model. This could also imply that the transition in the phase at energies of order 1~EeV suggested from observations may not be directly related to the transition from a Galactic to an extragalactic predominance, which may be taking place actually at lower energies, closer to the second knee. 

More precise measurements of the anisotropies in this energy range should give precious input for a better understanding of the different contributions to the CR fluxes and should also help to study the different processes originating them.

\section*{Acknowledgments}
We are grateful to I. Allekotte for useful comments and to D. Harari for discussions.
This work was supported by CONICET (PIP 2015-0369 and PIP 2021-0565).

\end{document}